\documentclass[12pt]{article}
\usepackage{amsmath}
\usepackage{epsf}
\usepackage{epsfig}
\usepackage{graphicx}
\usepackage[dvips]{lscape}
\usepackage{rotating}
\usepackage{subfigure}
\usepackage{subeqnarray}
\usepackage{setspace}
\usepackage{color}

\newcommand{\bR}{\mathbf{R}}

\def\bfX{{\bf X}}

\def\biota{\boldsymbol{\iota}}

\def\bGamma{\boldsymbol{\Gamma}}

\def\btheta{\boldsymbol{\theta}}
\def\bmu{\boldsymbol{\mu}}

\def\bomega{\boldsymbol{\omega}}

\newcommand{\E}{{\rm E}}

\newcommand{\Var}{{\rm Var}}
\newcommand{\SE}{{\rm SE}}
\newcommand{\bP}{\mathbf{P}}
\renewcommand{\P}{\textrm{P}}
\definecolor{RED}{rgb}{1,0,0}
\definecolor{BLUE}{rgb}{0,0,1}

\begin{document}

\title{~~
%
\\ [-1.25in]
{~~\\ \Huge\bf WARNING:  Physics Envy May Be \\
[.25in] Hazardous To Your Wealth!\thanks{The views and opinions expressed 
in this article are those of the authors only, and do not necessarily 
represent the views and opinions of AlphaSimplex Group, MIT, or any of 
their affiliates and employees. The authors make no representations or 
warranty, either expressed or implied, as to the accuracy or completeness 
of the information contained in this article, nor are they recommending 
that this article serve as the basis for any investment decision---this 
article is for information purposes only. Research support from 
AlphaSimplex Group and the MIT Laboratory for Financial Engineering is 
gratefully acknowledged.  We thank Jerry Chafkin, Peter Diamond, Arnout 
Eikeboom, Doyne Farmer, Gifford Fong, Jacob Goldfield, Tom Imbo, Jakub 
Jurek, Amir Khandani, Bob Lockner, Paul Mende, Robert Merton, Jun Pan, 
Roger Stein, Tina Vandersteel for helpful comments and discussion.}}}

\author{\Large Andrew W.~Lo\thanks{Harris \& Harris Group Professor, MIT Sloan
School of Management, and Chief Investment Strategist, AlphaSimplex Group,
LLC. Please direct all correspondence to: MIT Sloan School, 50 Memorial
Drive, E52--454, Cambridge, MA 02142--1347, {\tt alo@mit.edu} (email).}\ \
and Mark T.~Mueller\thanks{Senior Lecturer, MIT Sloan School of
Management, and Visiting Scientist, MIT Department of Physics, Center for
Theoretical Physics, 77 Massachusetts Avenue, Cambridge, MA 02142--1347,
{\tt mark.t.mueller@mac.com} (email).}}
\date{\Large This Draft: March 19, 2010 \\[.125in]}

\maketitle
\thispagestyle{empty}
\centerline{\large \bf Abstract}
\baselineskip 14pt
\vskip 20pt \noindent
The quantitative aspirations of economists and financial analysts have for 
many years been based on the belief that it should be possible to build 
models of economic systems---and financial markets in particular---that 
are as predictive as those in physics. While this perspective has led to a 
number of important breakthroughs in economics, ``physics envy'' has also 
created a false sense of mathematical precision in some cases. We 
speculate on the origins of physics envy, and then describe an alternate 
perspective of economic behavior based on a new taxonomy of uncertainty.  
We illustrate the relevance of this taxonomy with two concrete examples: 
the classical harmonic oscillator with some new twists that make physics 
look more like economics, and a quantitative equity market-neutral 
strategy. We conclude by offering a new interpretation of tail events, 
proposing an ``uncertainty checklist'' with which our taxonomy can be 
implemented, and considering the role that quants played in the current 
financial crisis. 

\newpage

\setcounter{page}{1} \pagenumbering{roman} \tableofcontents 
\newpage

\setcounter{page}{1}
\pagenumbering{arabic}
\onehalfspacing
\setlength{\belowdisplayskip}{0.250in}
\setlength{\abovedisplayskip}{0.250in}
\setlength{\belowdisplayshortskip}{0.250in}
\setlength{\abovedisplayshortskip}{0.250in}
\setcounter{equation}{0}
\setcounter{table}{0}
\setcounter{figure}{0}

\begin{center}
{\it Imagine how much harder physics would be if electrons had feelings!}
\end{center}
\hfill -- Richard Feynman, speaking at a Caltech graduation ceremony.

\section{Introduction}
\label{sec:intro}%
The Financial Crisis of 2007--2009 has re-invigorated the longstanding
debate regarding the effectiveness of quantitative methods in economics
and finance.  Are markets and investors driven primarily by fear and greed
that cannot be modeled, or is there a method to the market's madness that
can be understood through mathematical means?  Those who rail against the
quants and blame them for the crisis believe that market behavior cannot
be quantified and financial decisions are best left to individuals with
experience and discretion. Those who defend quants insist that markets are
efficient and the actions of arbitrageurs impose certain mathematical
relationships among prices that can be modeled, measured, and managed.  Is
finance a science or an art?

In this paper, we attempt to reconcile the two sides of this debate by 
taking a somewhat circuitous path through the sociology of economics and 
finance to trace the intellectual origins of this conflict---which we 
refer to as ``physics envy''---and show by way of example that ``the fault 
lies not in our models but in ourselves''.  By reflecting on the 
similarities and differences between economic phenomena and those of other 
scientific disciplines such as psychology and physics, we conclude that 
economic logic goes awry when we forget that human behavior is not nearly 
as stable and predictable as physical phenomena. However, this observation 
does not invalidate economic logic altogether, as some have argued. 

In particular, if, like other scientific endeavors, economics is an 
attempt to understand, predict, and control the unknown through 
quantitative analysis, the kind of uncertainty affecting economic 
interactions is critical in determining its successes and failures. 
Motivated by Knight's (1921) distinction between ``risk'' (randomness that 
can be fully captured by probability and statistics) and ``uncertainty'' 
(all other types of randomness), we propose a slightly finer 
taxonomy---fully reducible, partially reducible, and irreducible 
uncertainty---that can explain some of the key differences between finance 
and physics.  Fully reducible uncertainty is the kind of randomness that 
can be reduced to pure risk given sufficient data, computing power, and 
other resources.  Partially reducible uncertainty contains a component 
that can never be quantified, and irreducible uncertainty is the Knightian 
limit of unparametrizable randomness.  While these definitions may seem 
like minor extensions of Knight's clear-cut dichotomy, they underscore the 
fact that there is a continuum of randomness in between risk and 
uncertainty, and this nether region is the domain of economics and 
business practice. In fact, our taxonomy is reflected in the totality of 
human intellectual pursuits, which can be classified along a continuous 
spectrum according to the type of uncertainty involved, with religion at 
one extreme (irreducible uncertainty), economics and psychology in the 
middle (partially reducible uncertainty) and mathematics and physics at 
the other extreme (certainty). 

However, our more modest and practical goal is to provide a framework for 
investors, portfolio managers, regulators, and policymakers in which the 
efficacy and limitations of economics and finance can be more readily 
understood.  In fact, we hope to show through a series of examples drawn 
from both physics and finance that the failure of quantitative models in 
economics is almost always the result of a mismatch between the type of 
uncertainty in effect and the methods used to manage it.  Moreover, the 
process of scientific discovery may be viewed as the means by which we 
transition from one level of uncertainty to the next.  This framework can 
also be used to extrapolate the future of finance, the subject of this 
special volume of the {\it Journal of Investment Management}.  We propose 
that this future will inevitably involve refinements of the taxonomy of 
uncertainty and the development of more sophisticated methods for 
``full-spectrum'' risk management. 

We begin in Section \ref{sec:envy} by examining the intellectual milieu 
that established physics as the exemplar for economists, inevitably 
leading to the ``mathematization'' of economics and finance. The contrast 
and conflicts between physics and finance can be explained by considering 
the kinds of uncertainty they address, and we describe this taxonomy in 
Section \ref{sec:taxonomy}. We show how this taxonomy can be applied in 
two contexts in Sections \ref{sec:oscill} and \ref{sec:trading}, one drawn 
from physics (the harmonic oscillator) and the other drawn from finance (a 
quantitative trading strategy).  These examples suggest a new 
interpretation of so-called ``black swan'' events, which we describe in 
Section \ref{sec:swan}.  They also raise a number of practical issues that 
we address in Section \ref{sec:apply}, including the introduction of an 
``uncertainty checklist'' with which our taxonomy can be applied. Finally, 
in Section \ref{sec:quants} we turn to the role of quants in the current 
financial crisis, and consider three populist views that are either 
misinformed or based on incorrect claims, illustrating the benefits of a 
scientific approach to analysis crises. We conclude in Section 
\ref{sec:conclusion} with some speculation regarding the finance of the 
future. 

Before turning to these issues, we wish to specify the intended audience 
for this unorthodox and reflective article.  While we hope the novel 
perspective we propose and the illustrative examples we construct will 
hold some interest for our academic colleagues, this paper can hardly be 
classified as original research.  Instead, it is the engineer, research 
scientist, newly minted Wall Street quant, beleaguered investor, 
frustrated regulators and policymakers, and anyone else who cannot 
understand how quantitative models could have failed so spectacularly over 
the last few years that we intend to reach.  Our focus is not on the 
origins of the current financial crisis---there are now many popular and 
erudite accounts---but rather on developing a logical framework for 
understanding the role of quantitative models in theory and practice.  We 
acknowledge at the outset that this goal is ambitious, and beg the 
reader's indulgence as we attempt to reach our target audience through 
stylized examples and simplistic caricatures, rather than through formal 
theorem-and-proof.

\section{Physics Envy}
\label{sec:envy}
The fact that economics is still dominated by a single paradigm is a 
testament to the extraordinary achievements of one individual: Paul 
A.~Samuelson. In 1947, Samuelson published his Ph.D.~thesis titled 
\textit{Foundations of Economics Analysis}, which might have seemed 
presumptuous---especially coming from a Ph.D. candidate---were it not for 
the fact that it did, indeed, become the foundations of modern economic 
analysis. In contrast to much of the extant economic literature of the 
time, which was often based on relatively informal discourse and 
diagrammatic exposition, Samuelson developed a formal mathematical 
framework for economic analysis that could be applied to a number of 
seemingly unrelated contexts. Samuelson's (1947, p.~3) opening paragraph 
made his intention explicit (italics are Samuelson's): 
\begin{quote}
\baselineskip=14pt%
{\it The existence of analogies between central features of various
theories implies the existence of a general theory which underlies the
particular theories and unifies them with respect to those central
features}.  This fundamental principle of generalization by abstraction
was enunciated by the eminent American mathematician E.H.~Moore more than
thirty years ago.  It is the purpose of the pages that follow to work out
its implications for theoretical and applied economics.
\end{quote}
He then proceeded to build the infrastructure of what is now known as
microeconomics, routinely taught as the first graduate-level course in
every Ph.D.~program in economics today.  Along the way, Samuelson also
made major contributions to welfare economics, general equilibrium theory,
comparative static analysis, and business-cycle theory, all in a single
doctoral dissertation!

If there is a theme to Samuelson's thesis, it is the systematic
application of scientific principles to economic analysis, much like the
approach of modern physics.  This was no coincidence.  In Samuelson's
(1998, p.~1376) fascinating account of the intellectual origins of his
dissertation, he acknowledged the following:
\begin{quote}
\baselineskip=14pt%
Perhaps most relevant of all for the genesis of {\it Foundations}, Edwin
Bidwell Wilson (1879--1964) was at Harvard. Wilson was the great Willard
Gibbs's last (and, essentially only) prot\'eg\'e at Yale. He was a
mathematician, a mathematical physicist, a mathematical statistician, a
mathematical economist, a polymath who had done first-class work in many
fields of the natural and social sciences.  I was perhaps his only
disciple\ .\ .\ .\ I was vaccinated early to understand that economics and
physics could share the same formal mathematical theorems (Euler's theorem
on homogeneous functions, Weierstrass's theorems on constrained maxima,
Jacobi determinant identities underlying Le Chatelier reactions, etc.),
while still not resting on the same empirical foundations and certainties.
\end{quote}
\noindent Also, in a footnote to his statement of the general principle of
comparative static analysis, Samuelson (1947, p.~21) added, ``It may be
pointed out that this is essentially the method of thermodynamics, which
can be regarded as a purely deductive science based upon certain
postulates (notably the First and Second Laws of Thermodynamics)''. And
much of the economics and finance literature since {\it Foundations\/} has
followed Samuelson's lead in attempting to deduce implications from
certain postulates such as utility maximization, the absence of arbitrage,
or the equalization of supply and demand.  In fact, one of the most recent
milestones in economics---rational expectations---is founded on a single
postulate, around which a large and still-growing literature has
developed.

\subsection{The Mathematization of Economics and Finance}
Of course, the mathematization of economics and finance was not due to 
Samuelson alone, but was advanced by several other intellectual giants 
that created a renaissance of mathematical economics during the half 
century following the Second World War. One of these giants, Gerard 
Debreu, provides an eye-witness account of this remarkably fertile period: 
``Before the contemporary period of the past five decades, theoretical 
physics had been an inaccessible ideal toward which economic theory 
sometimes strove. During that period, this striving became a powerful 
stimulus in the mathematization of economic theory'' (Debreu, 1991, p.~2). 

What Debreu is referring to is a series of breakthroughs that not only
greatly expanded our understanding of economic theory, but also held out
the tantalizing possibility of practical applications involving fiscal and
monetary policy, financial stability, and central planning.  These
breakthroughs included:
\begin{itemize}\itemsep -0.0675in
\item Game theory (von Neumann and Morganstern, 1944; Nash, 1951)
\item General equilibrium theory (Debreu, 1959)
\item Economics of uncertainty (Arrow, 1964)
\item Long-term economic growth (Solow, 1956)
\item Portfolio theory and capital-asset pricing (Markowitz, 1954; Sharpe, 1964;
Tobin, 1958)
\item Option-pricing theory (Black and Scholes, 1973; Merton, 1973)
\item Macroeconometric models (Tinbergen, 1956; Klein, 1970)
\item Computable general equilibrium models (Scarf, 1973)
\item Rational expectations (Muth, 1961; Lucas, 1972)
\end{itemize}
Many of these contributions have been recognized by Nobel prizes, and they 
have permanently changed the field of economics from a branch of moral 
philosophy pursued by gentlemen scholars to a full-fledged scientific 
endeavor not unlike the deductive process with which Isaac Newton 
explained the motion of the planets from three simple laws.  Moreover, the 
emergence of econometrics, and the over-riding importance of theory in 
guiding empirical analysis in economics is similar to the tight 
relationship between experimental and theoretical physics.

The parallels between physics and finance are even closer, due to the fact 
that the Black-Scholes/Merton option-pricing formula is also the solution 
to the heat equation.  This is no accident, as Lo and Merton (2009) 
explain: 
\begin{quote}
\baselineskip=14pt%
The origins of modern financial economics can be traced to Louis 
Bachelier's magnificent dissertation, completed at the Sorbonne in 1900, 
on the theory of speculation.   This work marks the twin births of the 
continuous-time mathematics of stochastic processes and the 
continuous-time economics of option pricing.  In analyzing the problem of 
option pricing, Bachelier provides two different derivations of the 
Fourier partial differential equation as the equation for the probability 
density of what is now known as a Wiener process/Brownian motion.  In one 
of the derivations, he writes down what is now commonly called the 
Chapman-Kolmogorov convolution probability integral, which is surely among 
the earlier appearances of that integral in print.  In the other 
derivation, he takes the limit of a discrete-time binomial process to 
derive the continuous-time transition probabilities.  Along the way, 
Bachelier also developed essentially the method of images (reflection) to 
solve for the probability function of a diffusion process with an 
absorbing barrier.  This all took place five years before Einstein's 
discovery of these same equations in his famous mathematical theory of 
Brownian motion. 
\end{quote}
Not surprisingly, Samuelson was also instrumental in the birth of modern 
financial economics (see Samuelson, 2009), and---together with his 
Nobel-prize-winning protege Robert C.\ Merton---created much of what is 
now known as ``financial engineering'', as well as the analytical 
foundations of at least three multi-trillion-dollar industries 
(exchange-traded options markets, over-the-counter derivatives and 
structured products, and credit derivatives). 

The mathematization of neoclassical economics is now largely complete, 
with dynamic stochastic general equilibrium models, rational expectations, 
and sophisticated econometric techniques having replaced the less rigorous 
arguments of the previous generation of economists.  Moreover, the recent 
emergence of ``econophysics'' (Mantegna and Stanley, 2000)---a discipline 
that, curiously, has been defined not so much by its focus but more by the 
techniques (scaling arguments, power laws, and statistical mechanics) and 
occupations (physicists) of its practitioners---has only pushed the 
mathematization of economics and finance to new extremes.%
\footnote{However, this field is changing rapidly as physicists with 
significant practical experience in financial markets push the boundaries 
of theoretical and empirical finance; see Bouchaud, Farmer, and Lillo 
(2009) for an example of this new and exciting trend.} 

\subsection{Samuelson's Caveat}
Even as Samuelson wrote his remarkable {\it Foundations}, he was well 
aware of the limitations of a purely deductive approach.  In his 
introduction, he offered the following admonition (Samuelson, 1947, p.~3): 
\begin{quote}
\baselineskip=14pt%
.\ .\ .\ [O]nly the smallest fraction of economic writings, theoretical
and applied, has been concerned with the derivation of {\it operationally
meaningful\/} theorems.  In part at least this has been the result of the
bad methodological preconceptions that economic laws deduced from {\it a
priori\/} assumptions possessed rigor and validity independently of any
empirical human behavior. But only a very few economists have gone so far
as this.  The majority would have been glad to enunciate meaningful
theorems if any had occurred to them.  In fact, the literature abounds
with false generalization.

We do not have to dig deep to find examples.  Literally hundreds of
learned papers have been written on the subject of utility.  Take a little
bad psychology, add a dash of bad philosophy and ethics, and liberal
quantities of bad logic, and any economist can prove that the demand curve
for a commodity is negatively inclined.
\end{quote}

\noindent
This surprisingly wise and prescient passage is as germane today as it was 
over fifty years ago when it was first written, and all the more 
remarkable that it was penned by a twentysomething year-old graduate 
student.  The combination of analytical rigor and practical relevance was 
to become a hallmark of Samuelson's research throughout his career, and 
despite his theoretical bent, his command of industry practices and market 
dynamics was astonishing. Less gifted economists might have been able to 
employ similar mathematical tools and parrot his scientific perspective, 
but few would be able to match Samuelson's ability to distill the economic 
essence of a problem and then solve it as elegantly and completely. 

Unlike physics, in which pure mathematical logic can often yield useful
insights and intuition about physical phenomena, Samuelson's caveat
reminds us that a purely deductive approach may not always be appropriate
for economic analysis. As impressive as the achievements of modern physics
are, physical systems are inherently simpler and more stable than economic
systems, hence deduction based on a few fundamental postulates is likely
to be more successful in the former case than in the latter.  Conservation
laws, symmetry, and the isotropic nature of space are powerful ideas in
physics that simply do not have exact counterparts in economics because of
the nature of economic interactions and the types of uncertainty involved.

And yet economics is often the envy of the other social sciences, in which 
there are apparently even fewer unifying principles and operationally 
meaningful theorems.  Despite the well-known factions within economics, 
there is significant consensus among practicing economists surrounding the 
common framework of supply and demand, the principle of comparative 
advantage, the Law of One Price, income and substitution effects, net 
present value relations and the time value of money, externalities and the 
role of government, etc.  While false generalizations certainly abound 
among academics of all persuasions, economics does contain many true 
generalizations as well, and these successes highlight important 
commonalities between economics and the other sciences. 

Samuelson's genius was to be able to deduce operationally meaningful 
theorems despite the greater uncertainty of economic phenomena.  In this 
respect, perhaps the differences between physics and economics are not 
fundamental, but are due, instead, to the types of uncertainty inherent in 
the two respective disciplines.  We expand on this possibility in Sections 
\ref{sec:taxonomy}--\ref{sec:trading}.  Before turning to that framework, 
it is instructive to perform a side-by-side comparison of economics and 
its closest intellectual sibling---psychology. 

\subsection{Economics vs.\ Psychology}
The degree of physics envy among economists is more obvious when we
compare economics with the closely related field of psychology. Both
disciplines focus on human behavior, so one would expect them to have
developed along very similar ideological and methodological trajectories.
Instead, they have developed radically different cultures, approaching
human behavior in vastly different ways. Consider, first, some of the
defining characteristics of psychology:
\begin{itemize}\itemsep -0.0675in
\item Psychology is based primarily on observation and experimentation.
\item Field experiments are common.
\item Empirical analysis leads to new theories.
\item There are multiple theories of behavior.
\item Mutual consistency among theories is not critical.
\end{itemize}
Contrast these with the comparable characteristics of economics:
\begin{itemize}\itemsep -0.0675in
\item Economics is based primarily on theory and abstraction.
\item Field experiments are not common.
\item Theories lead to empirical analysis.
\item There are few theories of behavior.
\item Mutual consistency is highly prized.
\end{itemize}
Although there are, of course, exceptions to these generalizations, they
do capture much of the spirit of the two disciplines.%
\footnote{For example, there is a vast econometrics literature in which
empirical investigations are conducted, but almost always motivated by
theory and involving an hypothesis test of one sort or another.  For a
less impressionistic and more detailed comparison of psychology and
economics, See Rabin (1998, 2002).}
For example, while psychologists certainly do propose abstract theories of
human behavior from time to time, the vast majority of academic
psychologists conduct experiments. Although experimental economics has
made important inroads into the mainstream of economics and finance, the
top journals still publish only a small fraction of experimental papers,
the majority of publications consisting of more traditional theoretical
and empirical studies. Despite the fact that new theories of economic
behavior have been proposed from time to time, most graduate programs in
economics and finance teach only one such theory: expected utility theory
and rational expectations, and its corresponding extensions, e.g.,
portfolio optimization, the Capital Asset Pricing Model, and dynamic
stochastic general equilibrium models.  And it is only recently that
departures from this theory are not dismissed out of hand; less than a
decade ago, manuscripts containing models of financial markets with
arbitrage opportunities were routinely rejected from the top economics and
finance journals, in some cases without even being sent out to referees
for review.

But thanks to the burgeoning literature in behavioral economics and
finance, the Nobel prizes to Daniel Kahneman and Vernon Smith in 2002,
advances in the cognitive neurosciences, and the recent financial crisis,
psychological evidence is now taken more seriously by economists and
finance practitioners.  For example, going well beyond Keynes' (1936)
``animal spirits'', recent research in the cognitive neurosciences has
identified an important link between rationality in decision-making and
emotion,%
\footnote{See, for example, Grossberg and Gutowski (1987), Damasio (1994),
Elster (1998), Lo (1999), Lo and Repin (2002), Loewenstein (2000), and
Peters and Slovic (2000).}
implying that the two are not antithetical, but in fact complementary.  In
particular, emotions are the basis for a reward-and-punishment system that
facilitates the selection of advantageous behavior, providing a numeraire
for animals to engage in a ``cost-benefit analysis'' of the various
actions open to them (Rolls, 1999, Chapter 10.3). Even fear and
greed---the two most common culprits in the downfall of rational thinking,
according to most behavioralists---are the product of evolutionary forces,
adaptive traits that increase the probability of survival. From an
evolutionary perspective, emotion is a powerful tool for improving the
efficiency with which animals learn from their environment and their past.
When an individual's ability to experience emotion is eliminated, an
important feedback loop is severed and his decision-making process is
impaired.

These new findings imply that individual preferences and behavior may not
be stable through time, but are likely to be shaped by a number of
factors, both internal and external to the individual, i.e., factors
related to the individual's personality, and factors related to specific
environmental conditions in which the individual is currently situated.
When environmental conditions shift, we should expect behavior to change
in response, both through learning and, over time, through changes in
preferences via the forces of natural selection. These evolutionary
underpinnings are more than simple speculation in the context of financial
market participants.  The extraordinary degree of competitiveness of
global financial markets and the outsize rewards that accrue to the
``fittest'' traders suggest that Darwinian selection is at work in
determining the typical profile of the successful investor. After all,
unsuccessful market participants are eventually eliminated from the
population after suffering a certain level of losses.

This perspective suggests an alternative to the antiseptic world of 
rational expectations and efficient markets, one in which market forces 
and preferences interact to yield a much more dynamic economy driven by 
competition, natural selection, and the diversity of individual and 
institutional behavior.  This approach to financial markets, which we 
refer to as the ``Adaptive Markets Hypothesis'' (Farmer and Lo, 1999; 
Farmer, 2002; Lo, 2004, 2005; and Brennan and Lo, 2009), is a far cry from 
theoretical physics, and calls for a more sophisticated view of the role 
that uncertainty plays in quantitative models of economics and finance.  
We propose such a view in Section \ref{sec:taxonomy}. 

\section{A Taxonomy of Uncertainty}
\label{sec:taxonomy}
The distinctions between the various types of uncertainty are, in fact,
central to the differences between economics and physics.  Economists have
been aware of some of these distinctions for decades, beginning with the
University of Chicago economist Frank Knight's (1921) Ph.D.\ dissertation
in which he distinguished between two types of randomness: one that is
amenable to formal statistical analysis, which Knight called ``risk'', and
another that is not, which he called ``uncertainty''.  An example of the
former is the odds of winning at the roulette table, and an example of the
latter is the likelihood of peace in the Middle East within the next five
years.  Although Knight's motivation for making such a distinction is
different from ours---he was attempting to explain why some businesses
yield little profits (they take on risks, which easily become
commoditized) while others generate extraordinary returns (they take on
uncertainty)---nevertheless, it is a useful starting point for
understanding why physics seems so much more successful than economics. In
this section, we propose an even more refined taxonomy of uncertainty, one
capable of explaining the differences across the entire spectrum of
intellectual pursuits from physics to biology to economics to philosophy
and religion.

\subsection{Level 1: Complete Certainty}
\label{ssec:level1}
This is the realm of classical physics, an idealized deterministic world 
governed by Newton's laws of motion.  All past and future states of the 
system are determined exactly if initial conditions are fixed and 
known---nothing is uncertain.  Of course, even within physics, this 
perfectly predictable clockwork universe of Newton, Lagrange, LaPlace, and 
Hamilton was recognized to have limited validity as quantum mechanics 
emerged in the early twentieth century.  Even within classical physics, 
the realization that small perturbations in initial conditions can lead to 
large changes in the subsequent evolution of a dynamical system 
underscores how idealized and limited this level of description can be in 
the elusive search for truth. 

However, it must be acknowledged that much of the observable physical 
universe does, in fact, lie in this realm of certainty.  Newton's three 
laws explain a breathtakingly broad span of phenomena---from an apple 
falling from a tree to the orbits of planets and stars---and has done so 
in the same manner for more than 10 billion years. In this respect, 
physics has enjoyed a significant head start when compared to all the 
other sciences.

\subsection{Level 2: Risk without Uncertainty}
\label{ssec:level2}
This level of randomness is Knight's (1921) definition of risk: randomness
governed by a known probability distribution for a completely known set of
outcomes. At this level, probability theory is a useful analytical
framework for risk analysis. Indeed, the modern axiomatic foundations of
probability theory---due to Kolmogorov, Wiener, and others---is given
precisely in these terms, with a specified sample space and a specified
probability measure.  No statistical inference is needed, because we know
the relevant probability distributions exactly, and while we do not know
the outcome of any given wager, we know all the rules and the odds, and no
other information relevant to the outcome is hidden.  This is life in a
hypothetical honest casino, where the rules are transparent and always
followed.  This situation bears little resemblance to financial markets.

\subsection{Level 3: Fully Reducible Uncertainty}
\label{ssec:level3}
This is risk with a degree of uncertainty, an uncertainty due to unknown 
probabilities for a fully enumerated set of outcomes that we presume are 
still completely known.  At this level, classical (frequentist) 
statistical inference must be added to probability theory as an 
appropriate tool for analysis. By ``fully reducible uncertainty'', we are 
referring to situations in which randomness can be rendered arbitrarily 
close to Level-2 uncertainty with sufficiently large amounts of data using 
the tools of statistical analysis.  Fully reducible uncertainty is very 
much like an honest casino, but one in which the odds are not posted and 
must therefore be inferred from experience.  In broader terms, fully 
reducible uncertainty describes a world in which a single model generates 
all outcomes, and this model is parameterized by a finite number of 
unknown parameters that do not change over time and which can be estimated 
with an arbitrary degree of precision given enough data. 

The resemblance to the ``scientific method''---at least as it is taught in 
science classes today---is apparent at this level of uncertainty. One 
poses a question, develops a hypothesis, formulates a quantitative 
representation of the hypothesis (i.e., a model), gathers data, analyzes 
that data to estimate model parameters and errors, and draws a conclusion. 
Human interactions are often a good deal messier and more nonlinear, and 
we must entertain a different level of uncertainty before we encompass the 
domain of economics and finance.

\subsection{Level 4: Partially Reducible Uncertainty}
\label{ssec:level4}
Continuing our descent into the depths of the unknown, we reach a level of 
uncertainty that now begins to separate the physical and social sciences, 
both in philosophy and model-building objectives.  By Level-4 or 
``partially reducible'' uncertainty, we are referring to situations in 
which there is a limit to what we can deduce about the underlying 
phenomena generating the data.  Examples include data-generating processes 
that exhibit: (1)~stochastic or time-varying parameters that vary too 
frequently to be estimated accurately; (2)~nonlinearities too complex to 
be captured by existing models, techniques, and datasets; 
(3)~non-stationarities and non-ergodicities that render useless the Law of 
Large Numbers, Central Limit Theorem, and other methods of statistical 
inference and approximation; and (4)~the dependence on relevant but 
unknown and unknowable conditioning information. 

Although the laws of probability still operate at this level, there is a
non-trivial degree of uncertainty regarding the underlying structures
generating the data that cannot be reduced to Level-2 uncertainty, even
with an infinite amount of data.  Under partially reducible uncertainty,
we are in a casino that may or may not be honest, and the rules tend to
change from time to time without notice.  In this situation, classical
statistics may not be as useful as a Bayesian perspective, in which
probabilities are no longer tied to relative frequencies of repeated
trials, but now represent degrees of belief.  Using Bayesian methods, we
have a framework and lexicon with which partial knowledge, prior
information, and learning can be represented more formally.

Level-4 uncertainty involves ``model uncertainty'', not only in the sense 
that multiple models may be consistent with observation, but also in the 
deeper sense that more than one model may very well be generating the 
data. One example is a regime-switching model in which the data are 
generated by one of two possible probability distributions, and the 
mechanism that determines which of the two is operative at a given point 
in time is also stochastic, e.g., a two-state Markov process as in 
Hamilton (1989, 1990).  Of course, in principle, it is always possible to 
reduce model uncertainty to uncertainty surrounding the parameters of a 
single all-encompassing ``meta-model'', as in the case of a 
regime-switching process. Whether or not such a reductionist program is 
useful depends entirely on the complexity of the meta-model and nature of 
the application. 

At this level of uncertainty, modeling philosophies and objectives in
economics and finance begin to deviate significantly from those of the
physical sciences.  Physicists believe in the existence of fundamental
laws, either implicitly or explicitly, and this belief is often
accompanied by a reductionist philosophy that seeks the fewest and
simplest building blocks from which a single theory can be built. Even in
physics, this is an over-simplification, as one era's ``fundamental laws''
eventually reach the boundaries of their domains of validity, only to be
supplanted and encompassed by the next era's ``fundamental laws''.  The
classic example is, of course, Newtonian mechanics becoming a special case
of special relativity and quantum mechanics.

It is difficult to argue that economists should have the same faith in a 
fundamental and reductionist program for a description of financial 
markets (although such faith does persist in some, a manifestation of 
physics envy).  Markets are tools developed by humans for accomplishing 
certain tasks---not immutable laws of Nature---and are therefore subject 
to all the vicissitudes and frailties of human behavior. While behavioral 
regularities do exist, and can be captured to some degree by quantitative 
methods, they do not exhibit the same level of certainty and 
predictability as physical laws.  Accordingly, model-building in the 
social sciences should be much less informed by mathematical aesthetics, 
and much more by pragmatism in the face of partially reducible 
uncertainty.   We must resign ourselves to models with stochastic 
parameters or multiple regimes that may not embody universal truth, but 
are merely useful, i.e., they summarize some coarse-grained features of 
highly complex datasets.  

While physicists make such compromises routinely, they rarely need to 
venture down to Level 4, given the predictive power of the vast majority 
of their models.  In this respect, economics may have more in common with 
biology than physics.  As the great mathematician and physicist John von 
Neumann observed, ``If people do not believe that mathematics is simple, 
it is only because they do not realize how complicated life is''. 

\subsection{Level 5: Irreducible Uncertainty}
\label{ssec:level5}
Irreducible uncertainty is the polite term for a state of total ignorance; 
ignorance that cannot be remedied by collecting more data, using more 
sophisticated methods of statistical inference or more powerful computers, 
or thinking harder and smarter.  Such uncertainty is beyond the reach of 
probabilistic reasoning, statistical inference, and any meaningful 
quantification.  This type of uncertainty is the domain of philosophers 
and religious leaders, who focus on not only the unknown, but the 
unknowable. 

Stated in such stark terms, irreducible uncertainty seems more likely to
be the exception rather than the rule.  After all, what kinds of phenomena
are completely impervious to quantitative analysis, other than the deepest
theological conundrums?  The usefulness of this concept is precisely in
its extremity.  By defining a category of uncertainty that cannot be
reduced to any quantifiable risk---essentially an admission of
intellectual defeat---we force ourselves to stretch our imaginations to
their absolute limits before relegating any phenomenon to this level.

\subsection{Level $\infty$: Zen Uncertainty}
\label{ssec:levelinfinity}
Attempts to understand uncertainty are mere illusions; there is only
suffering.

\subsection{The Uncertainty Continuum}
As our sequential exposition of the five levels of uncertainty suggests,
whether or not it is possible to model economic interactions
quantitatively is not a black-and-white issue, but rather a continuum that
depends on the nature of the interactions.  In fact, a given phenomenon
may contain several levels of uncertainty at once, with some components
being completely certain and others irreducibly uncertain.  Moreover, each
component's categorization can vary over time as technology advances or as
our understanding of the phenomenon deepens.  For example, 3{,}000 years
ago solar eclipses were mysterious omens that would have been considered
Level-5 uncertainty, but today such events are well understood and can be
predicted with complete certainty (Level 1). Therefore, a successful
application of quantitative methods to modeling any phenomenon requires a
clear understanding of the level of uncertainty involved.

In fact, we propose that the failure of quantitative models in economics 
and finance is almost always attributable to a mismatch between the level 
of uncertainty and the methods used to model it.  In Sections 
\ref{sec:oscill}--\ref{sec:swan}, we provide concrete illustrations of 
this hypothesis.

\section{The Harmonic Oscillator}
\label{sec:oscill}
To illustrate the import of the hierarchy of uncertainty proposed in 
Section \ref{sec:taxonomy}, in this section we apply it to a well-known 
physical system: the simple harmonic oscillator.  While trivial from a 
physicist's perspective, and clearly meant to be an approximation to a 
much more complex reality, this basic model of high-school physics fits 
experimental data far better than even the most sophisticated models in 
the social sciences. Therefore, it is an ideal starting point for 
illustrating the differences and similarities between physics and finance. 
After reviewing the basic properties of the oscillator, we will inject 
certain types of noise into the system and explore the implications for 
quantitative models of its behavior as we proceed from Level 1 (Section 
\ref{ssec:osc1}) to Level 4 (Section \ref{ssec:osc4}).  As more noise is 
added, physics begins to look more like economics.%
\footnote{However, we must emphasize the illustrative nature of this 
example, and caution readers against too literal an interpretation of our 
use of the harmonic oscillator with noise.  We are not suggesting that 
such a model is empirically relevant to either physics or economics.  
Instead, we are merely exploiting its simplicity to highlight the 
differences between the two fields.}
We postpone a discussion of Level-5 uncertainty until Section 
\ref{sec:swan}, where we provide a more expansive perspective in which 
phenomena transition from Level 5 to Level 1 as we develop deeper 
understanding of their true nature.
 
\subsection{The Oscillator at Level 1}
\label{ssec:osc1}
\begin{figure}[htbp]
\centering
\includegraphics[scale=.6,clip,trim=50 500 100 -170]%
{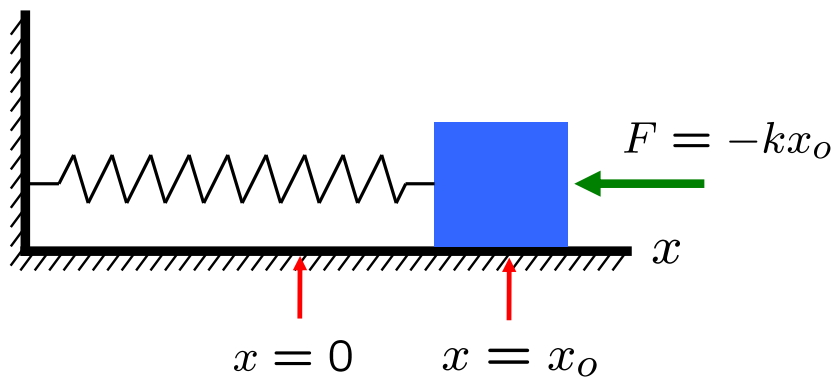}
\caption{Frictionless one-dimensional spring system.}
\label{fig:spring}
\end{figure}

Consider a mass situated at the origin $x\!=\!0$ of a frictionless 
surface, and suppose it is attached to a spring.  The standard model for 
capturing the behavior of this one-dimensional system when the mass is 
displaced from the origin---which dates back to 1660---is that the spring 
exerts a restoring force $F$ which is proportional to the displacement and 
in the opposite direction (see Figure \ref{fig:spring}); that is, $ F = 
-\,kx$ where $k$ is the spring constant and $x$ is the position of the 
mass $m$.  This hypothetical relation works so well in so many 
circumstances that physicists refer to it as a ``law'', Hooke's Law, in 
honor of its discoverer, Robert Hooke. 

By applying Newton's Second Law ($F=ma$) to Hooke's Law, we obtain
the following second-order linear differential equation:
\begin{equation}
\ddot x\ +\ \frac{k}{m}\,x\ \ =\ \ 0
\label{eq:ode}
\end{equation}
where $\ddot x$ denotes the second time-derivative of $x$.  The solution
to (\ref{eq:ode}) is well-known and given by:
\begin{equation}
x(t)\ \ =\ \ A\,\cos(\omega_o\,t + \phi)
\label{eq:odesol}
\end{equation}
where $\omega_o\equiv \sqrt{k/m}$ and $A$ and $\phi$ are constants that
depend on the initial conditions of the system.  This is the equation for
a harmonic oscillator with amplitude $A$, initial phase angle $\phi$,
period $T\!=\!2\pi\sqrt{m/k}$, and frequency $f\!=\!1/T$.  Despite its
deceptively pedestrian origins, the harmonic oscillator is ubiquitous in
physics, appearing in contexts from classical mechanics to quantum
mechanics to quantum field theory, and underpinning a surprisingly
expansive range of theoretical and applied physical phenomena.

\begin{figure}[htbp]
\centering
\includegraphics[scale=.5,clip,trim=0 0 0 0]{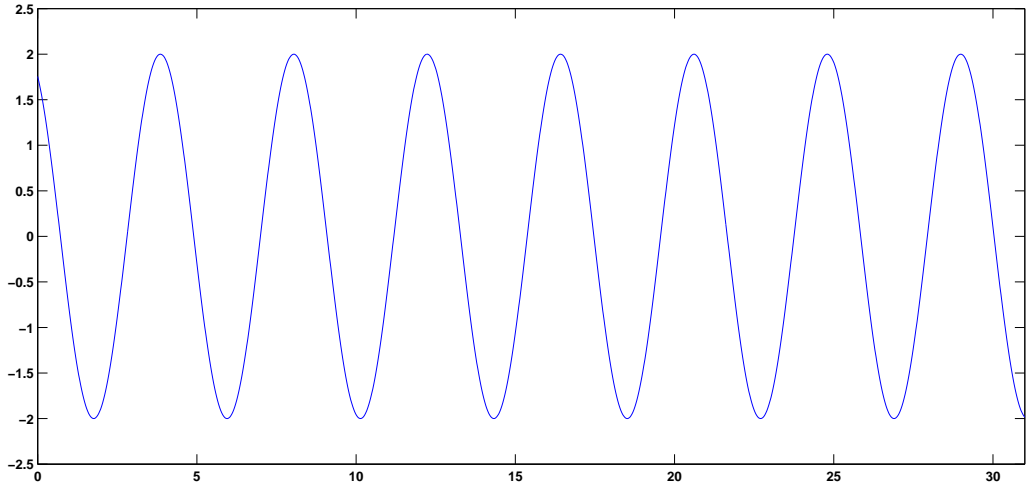}
\caption{Time series plot of the displacement $x(t)\!=\!A\cos(\omega_o\, t+\phi)$
of a harmonic oscillator with parameters $A\!=\!2$, $\omega_o\!=\!1.5$,
and $\phi\!=\!0.5$.}
\label{fig:coswave}
\end{figure}
At this stage, the model of the block's position (\ref{eq:odesol}) is
capturing a Level-1 phenomenon, perfect certainty.  For parameters
$A\!=\!2$, $\omega_o\!=\!1.5$, and $\phi\!=\!0.5$, Figure
\ref{fig:coswave} traces out the displacement of the block without
error---at time $t\!=\!3.5$, we know with certainty that $x\!=\ 1.7224$.

\subsection{The Oscillator at Level 2}
\label{ssec:osc2}
As satisfying as this knowledge is, physicists are the first to 
acknowledge that (\ref{eq:odesol}) is an idealization that is unlikely to 
be observed with real blocks and springs.  For example, surface and air 
friction will dampen the block's oscillations over time, and the initial 
displacement cannot exceed the spring's elastic limit otherwise the spring 
could break or be permanently deformed.  Such departures from the 
idealized world of (\ref{eq:odesol}) will cause us to leave the comforting 
world of Newtonian mechanics and Level-1 certainty. 

For the sake of exposition, consider the simplest departure from
(\ref{eq:odesol}), which is the introduction of an additive noise term to
the block's displacement:
\begin{equation}
x(t)\ \ =\ \ A\cos(\omega_o\,t+\phi)\ +\ \epsilon(t)~~,~~\epsilon(t)~{\rm IID}~
\mathcal{N}(0,\sigma^2_\epsilon)
\label{eq:noise}
\end{equation}
where `IID' stands for ``independently and identically distributed'', 
$\mathcal{N}(0,\sigma^2_\epsilon)$ indicates a normal distribution with 
mean 0 and variance $\sigma^2_\epsilon$, and all parameters, including 
$\sigma^2_\epsilon$, are fixed and known.%
\footnote{This example of Level-2 uncertainty corresponds to a number of 
actual physical phenomena, such as thermal noise in elements of an 
electronic circuit, where the analogue of the parameter $ 
\sigma^2_\epsilon $ is related in a fundamental way to the temperature of 
the system, described by an instance of a fluctuation-dissipation theorem, 
Nyquist's theorem (Reif, 1965, Chapter 15).}
Although $x(t)$ is no longer deterministic, Level-2 uncertainty implies 
that the probabilistic structure of the displacement is completely 
understood.  While we can no longer say with certainty that at time 
$t\!=\!3.5$, $x\!=\!1.7224$, we do know that if 
$\sigma_\epsilon\!=\!0.15$, the probability that $x$ falls outside the 
interval $[1.4284,2.0164]$ is precisely 5\%.

\subsection{The Oscillator at Level 3}
\label{ssec:osc3}
\begin{figure}[htbp]
\centering
\includegraphics[scale=.8,clip,trim=0 400 0 50]%
{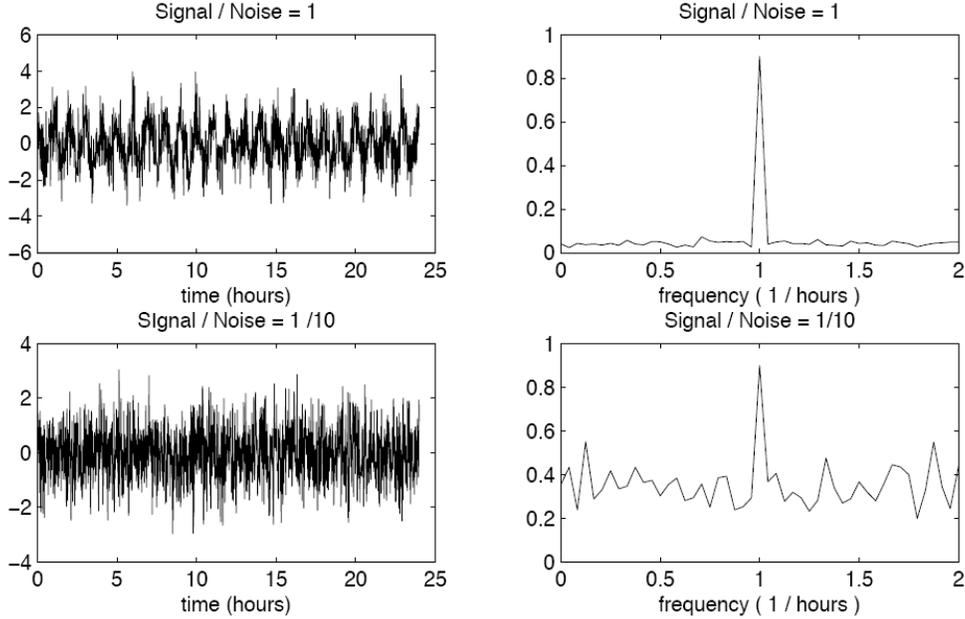}
\caption{Simulated time series and Fourier analysis of displacements
$x_t$ of a harmonic oscillator with a period of 1 hour, sampled at
1-minute intervals over a 24-hour period, with signal-to-noise
ratio of 1 (top graphs) and 0.10 (bottom graphs).}
\label{fig:sigtonoise}
\end{figure}

Level 3 of our taxonomy of uncertainty is fully-reducible uncertainty,
which we have defined as uncertainty that can be reduced arbitrarily
closely to pure risk (Level 2) given a sufficient amount of data. Implicit
in this definition is the assumption that the future is exactly like the
past in terms of the stochastic properties of the system, i.e.,
stationarity.%
\footnote{In physics, the stationarity of physical laws is often taken for 
granted.  For example, fundamental relationships such as Newton's 
$F\!=\!ma$ have been valid over cosmological time scales, and appear to be 
accurate descriptions of many physical phenomena from very early in the 
history of the Universe (Weinberg, 1977). Economics boasts no such 
equivalents.  However, other physical systems exhibit localized 
nonstationarities that fall outside Level-3 uncertainty (e.g., the 
properties of semiconductors as they are subjected to time-varying and 
non-stationary temperature fluctuations, vibration, and electromagnetic 
radiation), and entire branches of physics are devoted to such 
non-stationarities, e.g., non-equilibrium statistical mechanics.} 
In the case of the harmonic oscillator, this assumption means that the
oscillator (\ref{eq:odesol}) is still the data-generating process with
fixed parameters, but the parameters and the noise distribution are
unknown:
\begin{equation}
x(t)\ \ =\ \ A\cos(\omega_o\,t+\phi)\ +\ \epsilon(t)~~,~~
\E[\epsilon(t)]=0~~,~~\E[\epsilon(t_1)\epsilon(t_2)]=0
\ \forall~t_1\neq t_2\ .
\label{eq:noise2}
\end{equation}
However, under Level-3 uncertainty, we can estimate all the relevant
parameters arbitrarily accurately with enough data, which is a trivial
signal-processing exercise.  In particular, although the timescale and the
units in which the amplitude is measured in (\ref{eq:noise2}) are
completely arbitrary, for concreteness let the oscillator have a period of
1 hour given a 24-hour time sample period.  With data sampled every
minute, this system yields a sample size
of 1{,}440 observations.%
\footnote{Alternatively, for those who may wish to picture a system with a
longer time scale, this example can be interpreted as sampling daily
observations of the oscillator with a period of 60 days over a total time
period of slightly less that four years.}
Assuming a signal-to-noise ratio of $0.1$ and applying the Fast Fourier 
Transform (FFT) to this time series yields an excellent estimate of the 
oscillator's frequency, as Figure \ref{fig:sigtonoise} demonstrates. If 
this system were even a coarse approximation to business cycles and stock 
market fluctuations, economic forecasting would be a trivial task. 

\subsection{The Oscillator at Level 4}
\label{ssec:osc4}
We now turn to Level-4 uncertainty, the taxon in which uncertainty is only 
partially reducible.  To illustrate the characteristics of this level, 
suppose that the displacement $x(t)$ is not governed by a single 
oscillator with additive noise, but is subject to regime shifts between 
two oscillators without noise (we will consider the case of an oscillator 
with additive noise below). Specifically, let: 
\begin{subeqnarray}
x(t) &=& I(t)\,x_1(t)\ +\ \bigl(1-I(t)\bigr)\,x_2(t) \label{eq:xregime}\\
x_i(t) &=& A_i\cos(\omega_i\,t+\phi_i)~~,~~i=1,2
\end{subeqnarray}
where the binary indicator $I(t)$ determines which of the two oscillators $x_1(t)$ or $x_2(t)$ is generating the observed process $x(t)$, and let $I(t)$ be a simple two-state Markov process with the following simple transition probability matrix $\bP$:
\begin{equation}
\bP \ \ \equiv\ \ \bordermatrix{ & I(t)=1& I(t)=0 \cr
 I(t-1)=1 & 1-p & p \cr
 I(t-1)=0 & p & 1-p \cr}\ .
 \label{eq:chain}
\end{equation}
As before, we will assume that we observe the system once per minute over
a timespan of 24 hours, and let the two oscillators' periods be 30 and 60
minutes, respectively; their dimensionless ratio is independent of the
chosen timescale, and has been selected arbitrarily. Although the most
general transition matrix could allow for different probabilities of
exiting each state, for simplicity we have taken these probabilities to be
equal to a common value $ p $. The value of $p$ determines the half-life
of remaining in a given state, and we will explore the effects of varying
$p$.  We also impose restrictions on the parameters of the two oscillators
so that the observed displacement $x(t)$ and its first time-derivative
$\dot{x}(t)$ are continuous across regime switches, which ensures that the
observed time series do not exhibit
sample-path discontinuities or ``structural breaks''.%
\footnote{Specifically, we require that at any regime-switching time, the
block's position and velocity are continuous. Of course the block's second
time-derivative, i.e., its acceleration, will not be continuous across
switching times.  Therefore, in transiting from regime $i$ with angular
frequency $\omega_i$ to regime $j$ with angular frequency $\omega_j$, we
require that $A_i\cos( \omega_i \Delta t_i + \phi_i )  = A_j \cos( \phi_j)
$ and $ A_i \omega_i \sin( \omega_i \Delta t_i + \phi_i )  = A_j \omega_j
\sin( \phi_j )$, where $\Delta t_i$ is the time the system has just spent
in regime $i$, which is known at the time of the transition.  These
equations have a unique solution for $A_j$ and $\phi_j$ in terms of $A_i$
and $\phi_i$:
\[
A_j^2 = A_i^2 [ \cos^2( \omega_i \Delta t_i + \phi_i) +
(\omega_i / \omega_j)^2  \sin^2( \omega_i \Delta t_i + \phi_i) ] ~~,~~
\tan( \phi_j ) =  (\omega_i / \omega_j)\tan( \omega_i \Delta t_i + \phi_i )
\]
In simulating $x(t)$, these equations are applied iteratively, starting
from the chosen initial conditions at the beginning of the sample, and the
regime shifts are governed by the Markov chain (\ref{eq:chain}).

It is also worth mentioning that energy is not conserved in this system,
since the switching of frequencies can be characterized as some external
influence injecting or extracting energy from the oscillator.
Accordingly, the amplitude of the observed time series can and does
change, and the system is unstable in the long run.}
This example of Level-4 uncertainty illustrates one possible form of the 
stochastic behavior of the oscillator frequency, namely a regime switching 
model.  Such stochastic oscillators can be useful in building models of a 
variety of physical systems (see Gitterman, 2005). For example, the 
description of wave propagation in random media involves the behavior of 
certain types of stochastic oscillators, as in the propagation of sound 
waves through the ocean. 

\begin{figure}[htbp]
\centering
\includegraphics[scale=.6,clip,trim=0 50 0 0]{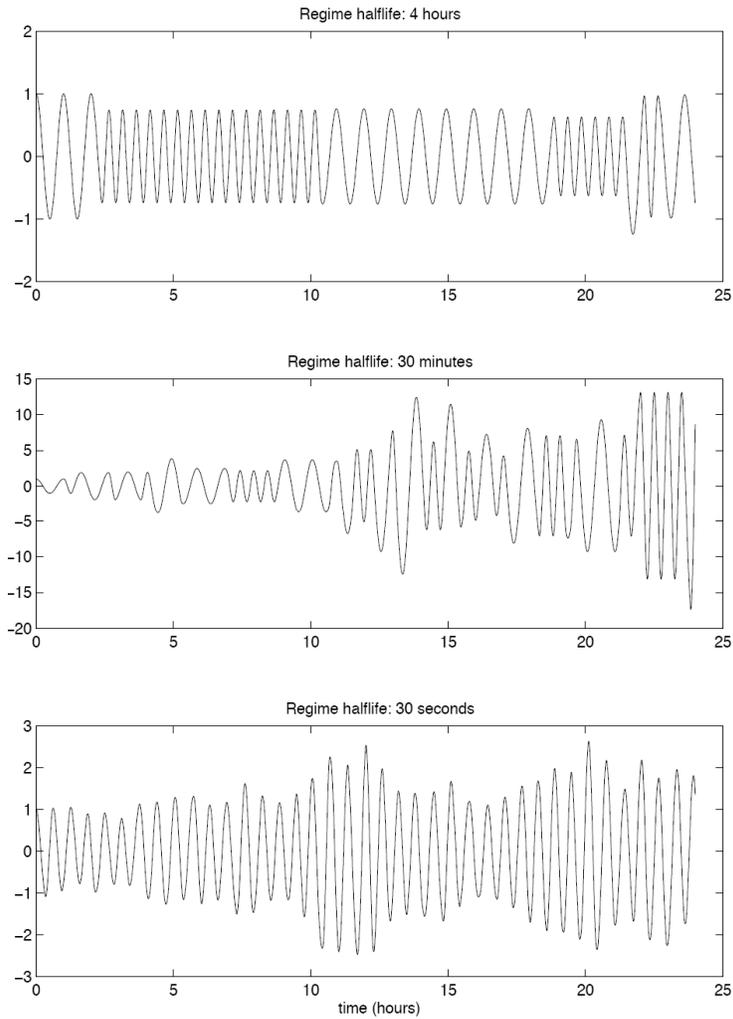}
\caption{Simulated time series of a harmonic oscillator with regime-switching
parameters in which the probability of a regime switch is calibrated to
yield a regime half-life of 4 hours (top panel), 30 minutes (middle
panel), and  30 seconds (bottom panel).  The oscillator's parameters are
completely specified by the half-life (or equivalently the transition
probability) and the frequencies in each regime.}
\label{fig:oscill2}
\end{figure}

In keeping with our intention to illustrate Level-4 uncertainty, suppose 
that none of the details of the structure of $x(t)$ are known to the 
observer attempting to model this system.  Figure \ref{fig:oscill2} shows 
three different examples of the observed time series $x(t)$ generated by 
(\ref{eq:xregime}). Each panel corresponds to a different value for the 
switching probability $p$: the top panel corresponds to a half-life of 4 
hours (significantly greater than the periods of either oscillator); the 
middle panel corresponds to a half-life of 30 minutes (the same period as 
oscillator 1 and comparable to oscillator 2); and the bottom panel 
corresponds to a half-life of 30 seconds (considerably shorter than the 
period of both oscillators). 

\begin{figure}[htbp]
\centering
\includegraphics[scale=.6,clip,trim=0 50 0 0]{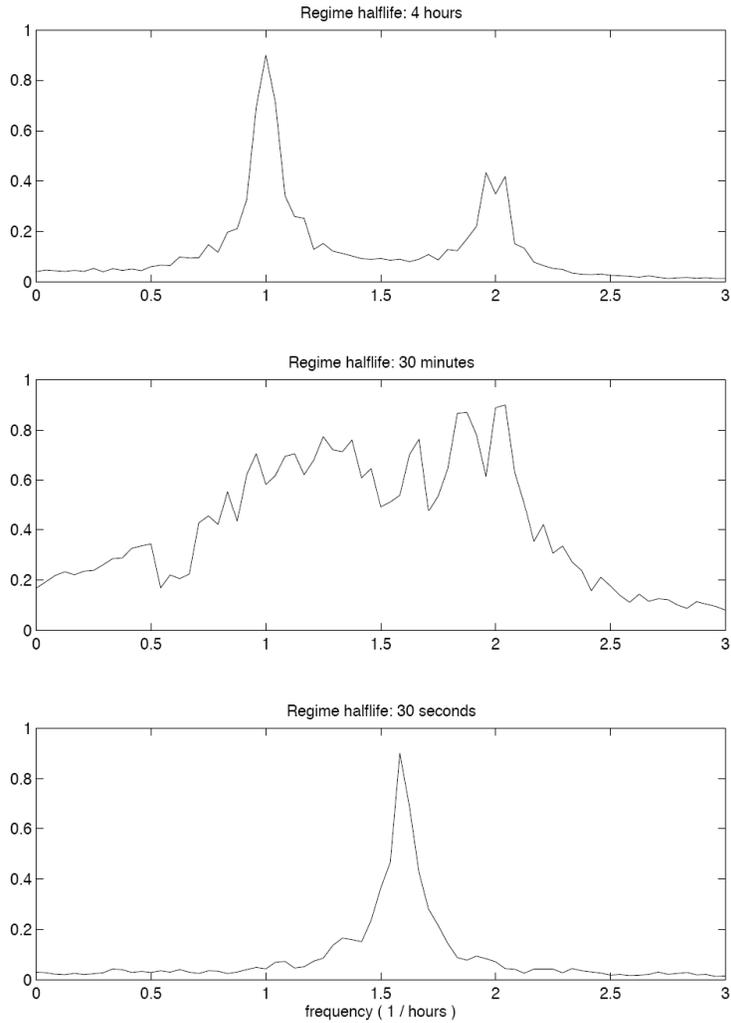}
\caption{Fast Fourier transforms of the simulated time series of a harmonic 
oscillator with regime-switching parameters in which the probability of a 
regime switch is calibrated to yield a regime half-life of 4 hours (top 
panel), 30 minutes (middle panel), and  30 seconds (bottom panel). The 
oscillator's parameters are completely specified by the half-life (or 
equivalently the transition probability) and the frequencies in each 
regime.} \label{fig:regime1} 
\end{figure}

In these cases, the behavior of $x(t)$ is less transparent than in the
single-oscillator case.  However, in the two extreme cases where the
values of $p$ imply either a much greater or much smaller period than the
two oscillators, there is hope. When the regime half-life is much greater
than both oscillators' periods, enough data can be collected during each
of the two regimes to discriminate between them.  In fact, visual
inspection of the Fourier transform depicted in the top panel of Figure
\ref{fig:regime1} confirms this intuition. At the other extreme, when the
regime half-life is much shorter than the two oscillators' periods, the
system behaves as if it were a single oscillator with a frequency equal to
the harmonic mean of the two oscillators' frequencies, and with an
amplitude that is stochastically modulated. The Fourier transform in the
bottom panel of Figure \ref{fig:regime1} confirms that a single effective
frequency is present.  Despite the fact that two oscillators are, in fact,
generating the data, for all intents and purposes, modeling this system as
a single oscillator will yield an excellent approximation.

The most challenging case is depicted in the middle panel of Figure
\ref{fig:regime1}, which corresponds to a value of $p$ that implies a
half-life comparable to the two oscillators' periods. Neither the time
series nor its Fourier transform offers any clear indication as to what is
generating the data.  And recall that these results do not yet reflect the
effects of any additive noise terms as in (\ref{eq:noise2}).

\begin{figure}[htbp]
\centering
\includegraphics[scale=.6,clip,trim=0 50 0 0]{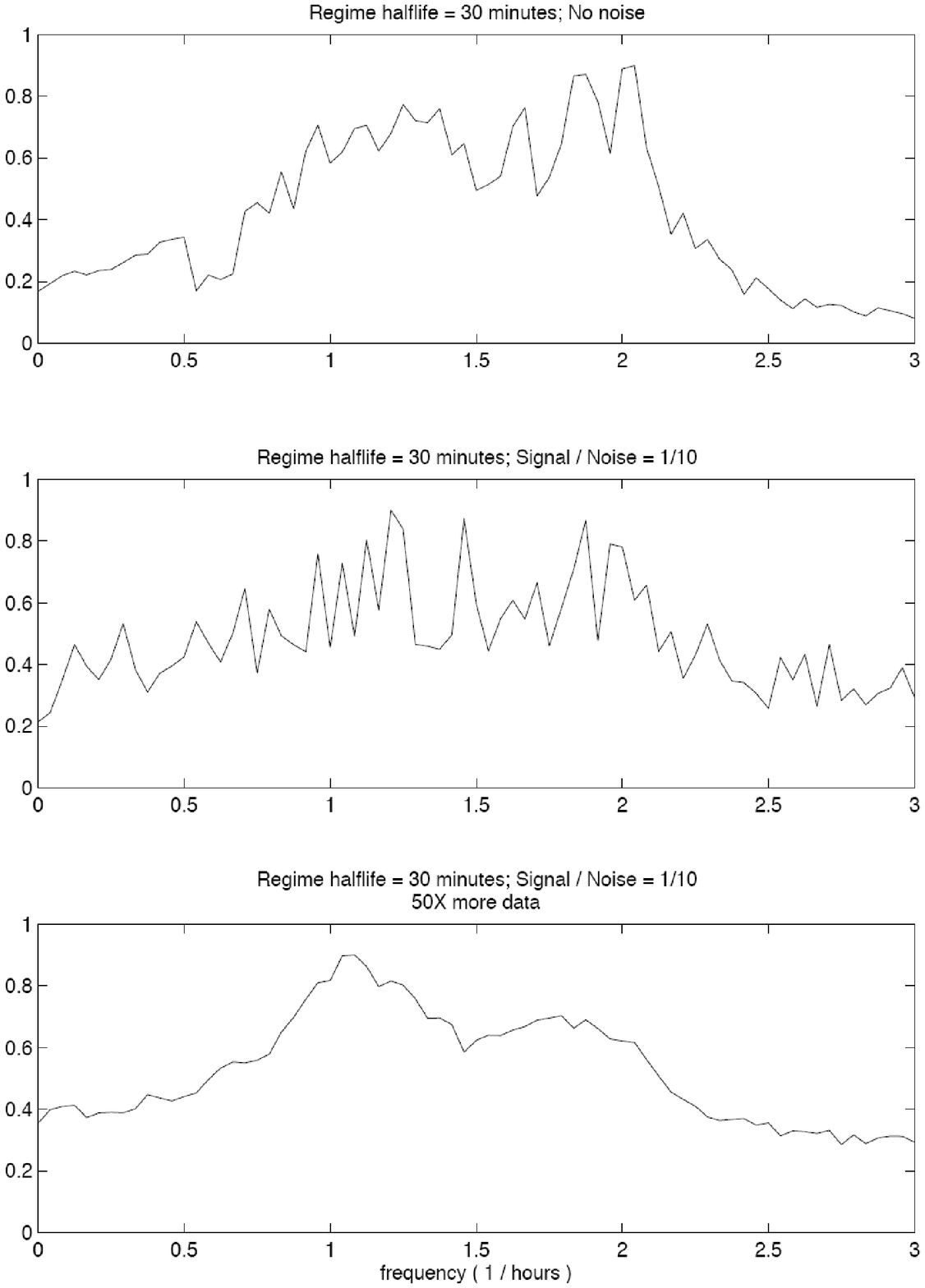}
\caption{Fast Fourier transforms of simulated time series the two-state
regime-switching harmonic oscillator with additive noise. For comparison,
the case without additive noise and where the regime half-life is comparable
to the oscillators' periods is given in the top panel.  The middle panel
corresponds to the case with additive noise and regime half-life comparable
to the oscillators' periods, and the bottom panel is the same analysis applied
to a dataset 50 times longer than the other two panels.}
\label{fig:regime2}
\end{figure}

If we add Gaussian white noise to (\ref{eq:xregime}), the
signal-extraction process becomes even more challenging, as Figure
\ref{fig:regime2} illustrates.  In the top panel, we reproduce the Fourier
transform of the middle panel of Figure \ref{fig:regime1}  (that is, with
no noise), and in the two panels below, we present the Fourier transforms
of the same regime-switching model with additive Gaussian noise,
calibrated with the same parameters as in the middle panel of Figure
\ref{fig:regime1} for the oscillators, and with the additive noise
component calibrated to yield a signal-to-noise ratio of $0.1$.  The
middle panel of Figure \ref{fig:regime2} is the Fourier transform of this
new simulation using the same sample size as in Figure \ref{fig:regime1},
and the bottom panel contains the Fourier transform applied to a dataset
50 times larger. There is no discernible regularity in the oscillatory
behavior of the middle panel, and only when we use significantly more data
do the two characteristic frequencies begin to emerge.

\begin{figure}[htbp]
\centering
\includegraphics[scale=.6,clip,trim=0 50 0 0]{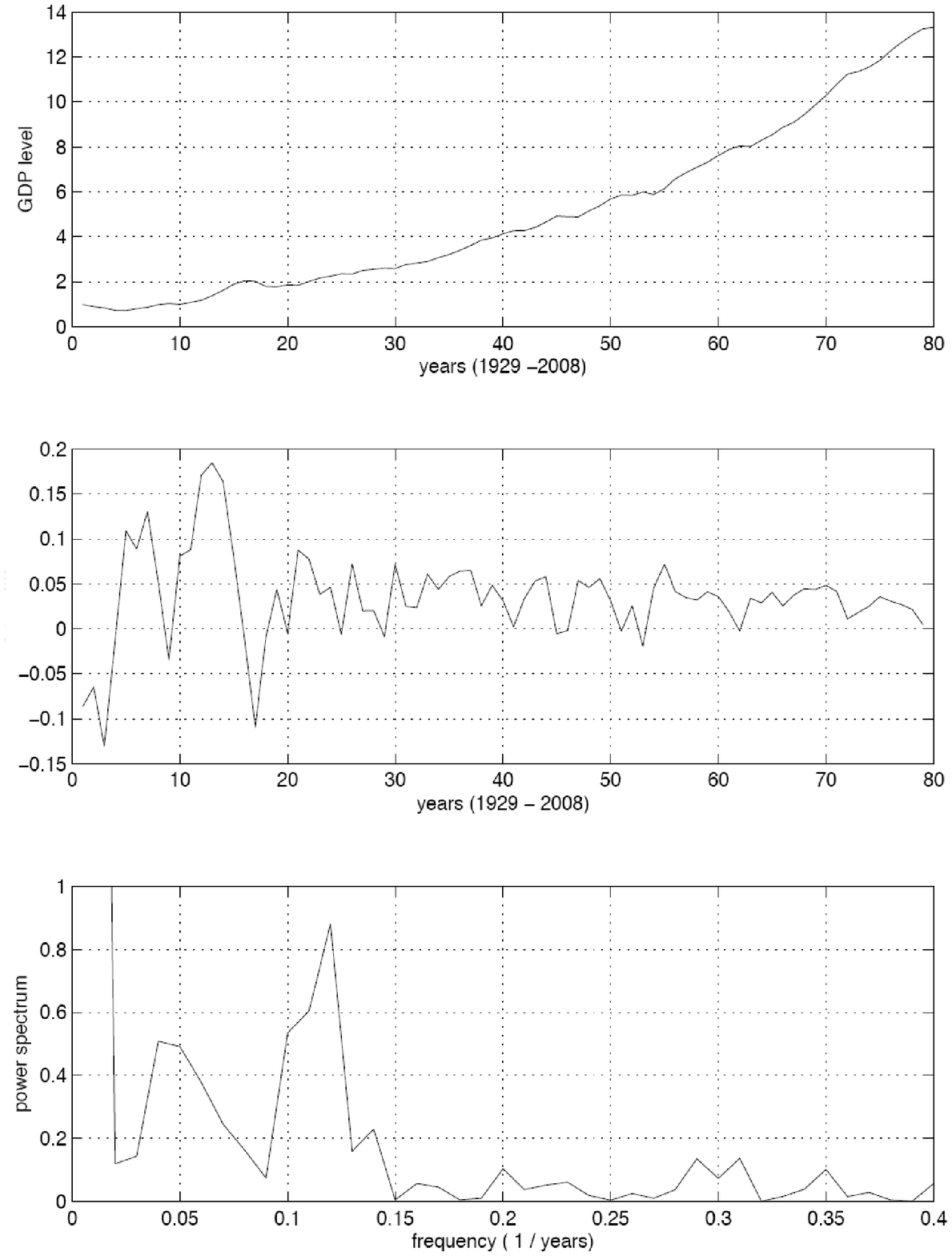}
\caption{Annual levels (top panel), growth rates (middle panel),
and Fourier transform of U.S.\ GDP (bottom panel) in 2005 dollars,
from 1929 to 2008.}
\label{fig:GDP}
\end{figure}

The relevance of Level-4 uncertainty for economics is obvious when we 
compare the middle panel of Figure \ref{fig:regime2} to the Fourier 
transform of a standard economic time series such as growth rates for 
U.S.\ real gross domestic product from 1929 to 2008, displayed in the 
bottom panel of Figure \ref{fig:GDP}.  The similarities are striking, 
which may be why Paul Samuelson often quipped that ``economists have 
predicted five out of the past three recessions''. 

Note the markedly different data requirements between this model with 
Level-4 uncertainty and those with Level-3 uncertainty.  Considerably more 
data is needed just to identify the presence of distinct regimes, and this 
is a necessary but not sufficient condition for accurate estimation of the 
model's parameters, i.e., the oscillator frequencies, the transition 
matrix for the Markov process, and the variance of the additive noise. 
Moreover, even if we are able to obtain sufficient data to accurately 
estimate the model's parameters, it may still be impossible to construct 
reliable forecasts if we cannot identify in real time which regime we 
currently inhabit.  In the financial markets context, such a model implies 
that even if markets are inefficient---in the sense that securities 
returns are not pure random walks---we may still be unable to profit from 
it, a particularly irksome state of affairs that might be called the 
``Malicious Markets Hypothesis''. 

If Tycho, Kepler, Galileo and Newton had been confronted with such 
``quasi-periodic'' data for the cyclical paths of the planets, it may have 
taken centuries, not years, of observational data for them to arrive at 
the same level of understanding of planetary motion that we have today. 
Fortunately, as challenging as physics is, much of it apparently does not 
operate at this level of uncertainty. Unfortunately, financial markets can 
sometimes operate at an even deeper level of uncertainty, uncertainty that 
cannot be reduced to any of the preceding levels simply with additional 
data.  We turn to this level in Section \ref{sec:swan}, but before doing 
so, in the next section we provide an application of our taxonomy of 
uncertainty to a financial context. 

\section{A Quantitative Trading Strategy}
\label{sec:trading}
Since the primary focus of this paper is the limitations of a 
physical-sciences approach to financial applications, in this section we 
apply our taxonomy of uncertainty to the mean-reversion strategy of 
Lehmann (1990) and Lo and MacKinlay (1990a), in which a portfolio of 
stocks is constructed by buying previous under-performing stocks, and 
short-selling previous outperforming stocks by the same dollar amount. 
This quantitative equity market-neutral strategy is simple enough to yield 
analytically tractable expressions for its statistical properties, and 
realistic enough to illustrate many of the practical challenges of this 
particular part of the financial industry, which has come to be known as 
``statistical arbitrage'' or ``statarb''.  This strategy has been used 
more recently by Khandani and Lo (2007, 2008) to study the events 
surrounding the August 2007 ``Quant Meltdown'', and the exposition in this 
section closely follows theirs.  To parallel our sequential exposition of 
the oscillator in Section \ref{sec:oscill}, we begin in Section 
\ref{ssec:statarb_level1} with those few aspects of statarb that may be 
interpreted as Level-1 certainty, and focus on progressively more 
uncertain aspects of the strategy in Sections 
\ref{ssec:statarb_level2}--\ref{ssec:statarb_level4}.  As before, we will 
postpone any discussion of Level-5 uncertainty to Section \ref{sec:swan} 
where we address the full range of uncertainty more directly.

\subsection{StatArb at Level 1}
\label{ssec:statarb_level1}%
Perhaps the most obvious difference between physics and financial 
economics is the fact that almost no practically relevant financial 
phenomena falls into the category of Level 1, perfect certainty.  The very 
essence of finance is the interaction between uncertainty and investment 
decisions, hence in a world of perfect certainty, financial economics 
reduces to standard microeconomics.  Unlike the oscillator of Section 
\ref{sec:oscill}, there is no simple deterministic model of financial 
markets that serves as a useful starting point for practical 
considerations, except perhaps for simple default-free bond-pricing 
formulas and basic interest-rate calculations.  For example, consider a 
simple deterministic model of the price $P_t$ of a financial security: 
\begin{equation}
P_t\ \ =\ \ P_{t-1}\ +\ X_t
\label{eq:certainty}
\end{equation}
where $X_t$ is some known increment.  If $X_t$ is known at date $t\!-\!1$, 
then a positive value will cause investors to purchase as many shares of 
the security as possible in anticipation of the price appreciation, and a 
negative value will cause investors to short-sell as many shares as 
possible to profit from the certain price decline.  Such behavior implies 
that $P_t$ will take on only one of two extreme values---0 or 
$\infty$---which is clearly unrealistic.  Any information regarding future 
price movements will be exploited to the fullest extent possible, hence 
deterministic models like (\ref{eq:certainty}) are virtually useless in 
financial contexts except for the most elementary pedagogical purposes. 

\subsection{StatArb at Level 2}
\label{ssec:statarb_level2}%
Consider a collection of $N$ securities and denote by $\bR_t$ the $N
{\times} 1$-vector of their date-$t$ returns $[ R_{1t} \cdots R_{Nt} ]'$.
To be able to derive the statistical properties of any strategy based on
these security returns, we require the following assumption:
\begin{itemize}
\item [\textbf{(A1)}] $\bR_t$ is a jointly covariance-stationary 
multivariate stochastic process with known distribution, which we take 
to be Gaussian with expectation and autocovariances: 
\begin{eqnarray*}
{\rm E}[\bR_t] &\equiv & \bmu\ \ =\ \ [ \mu_1\ \mu_2\ \cdots\ \mu_N ]' \\
{\rm E}[(\bR_{t-k}-\bmu)(\bR_t-\bmu)'] &\equiv & \bGamma_k
\end{eqnarray*}
where, with no loss of generality, we let $k \geq 0$ since $\bGamma_k 
= \bGamma'{}_{-k}$. 
\end{itemize}
This assumption is the embodiment of Level-2 uncertainty where the 
probability distributions of all the relevant variables are well-defined, 
well-behaved, and stable over time.%
\footnote{The assumption of multivariate normality in (A1) is not strictly 
necessary for the results in this section, but is often implicitly 
assumed, e.g., in Value-at-Risk computations of this portfolio's return, 
in the exclusive focus on the first two moments of return distributions, 
and in assessing the  statistical significance of associated regression 
diagnostics.}

Given these $N$ securities, consider a long/short market-neutral equity
strategy consisting of an equal dollar amount of long and short positions,
where at each rebalancing interval, the long positions consist of
``losers'' (underperforming stocks, relative to some market average) and
the short positions consist of ``winners'' (outperforming stocks, relative
to the same market average).  Specifically, if $\omega_{it}$ is the
portfolio weight of security $i$ at date $t$, then
\begin{eqnarray}
\omega_{it}(k) &=& -\,\frac{1}{N}(R_{it-k}-R_{mt-k})~~,~~ R_{mt-k}\ \equiv\
\frac{1}{N}\sum_{i=1}^N R_{it-k}
\label{eq:contrarian}
\end{eqnarray}
for some $k\!>\!0$.  By buying the previous losers and selling the
previous winners at each date, such a strategy actively bets on mean
reversion across all $N$ stocks, profiting from reversals that occur
within the rebalancing interval.%
\footnote{However, Lo and MacKinlay (1990a) show that this need not be the
only reason that contrarian investment strategies are profitable. In
particular, if returns are positively cross-autocorrelated, they show that
a return-reversal strategy will yield positive profits on average, even if
individual security returns are serially independent.  The presence of
stock market overreaction, i.e., negatively autocorrelated individual
returns, enhances the profitability of the return-reversal strategy, but
is not required for such a strategy to earn positive expected returns.}
For this reason, (\ref{eq:contrarian}) has been called a ``contrarian'' 
trading strategy that benefits from overreaction, i.e., when 
underperformance is followed by positive returns and vice-versa for 
outperformance.  Also, since the portfolio weights are proportional to the 
differences between the market index and the returns, securities that 
deviate more positively from the market at time $t\!-\!k$ will have 
greater negative weight in the date-$t$ portfolio, and vice-versa.  Also, 
observe that the portfolio weights are the negative of the degree of 
outperformance $k$ periods ago, so each value of $k$ yields a somewhat 
different strategy. Lo and MacKinlay (1990a) provide a detailed analysis 
of the unleveraged returns (\ref{eq:return}) of the contrarian trading 
strategy, tracing its profitability to mean reversion in individual stock 
returns as well as positive lead/lag effects and cross-autocorrelations 
across stocks and across time. 

Note that the weights (\ref{eq:contrarian}) have the property that they
sum to 0, hence (\ref{eq:contrarian}) is an example of an ``arbitrage'' or
``market-neutral'' portfolio where the long positions
are exactly offset by the short positions.%
\footnote{Such a strategy is more accurately described as a
``dollar-neutral'' portfolio since dollar-neutral does not necessarily
imply that a strategy is also market-neutral.  For example, if a portfolio
is long \$100MM of high-beta stocks and short \$100MM of low-beta stocks,
it will be dollar-neutral but will have positive market-beta exposure.  In
practice, most dollar-neutral equity portfolios are also constructed to be
market-neutral, hence the two terms are used almost interchangeably.}
As a result, the portfolio ``return'' cannot be computed in the standard
way because there is no net investment. In practice, however, the return
of such a strategy over any finite interval is easily calculated as the
profit-and-loss of that strategy's positions over the interval divided by
the initial capital required to support those positions.  For example,
suppose that a portfolio consisting of \$100MM of long positions and
\$100MM of short positions generated profits of \$2MM over a one-day
interval.  The return of this strategy is simply \$2MM divided by the
required amount of capital to support the \$100MM long/short positions.
Under Regulation T, the minimum amount of capital required is \$100MM
(often stated as $2\!:\!1$ leverage, or a 50\% margin requirement), hence
the return to the strategy is 2\%. If, however, the portfolio manager is a
broker-dealer, then Regulation T does not apply (other regulations govern
the capital adequacy of broker-dealers), and higher levels of leverage may
be employed. For example, under certain conditions, it is possible to
support a \$100MM long/short portfolio with only \$25MM of
capital---leverage ratio of $8\!:\!1$---which implies a portfolio
return of $\$2/\$25\!=\!8\%$.%
\footnote{The technical definition of leverage---and the one used by the
U.S.\ Federal Reserve, which is responsible for setting leverage
constraints for broker-dealers---is given by the sum of the absolute
values of the long and short positions divided by the capital, so:
\[
\frac{~|\$100| + |-\!\$100|~}{\$25}\ \ =\ \ 8\ .
\]
}
Accordingly, the gross dollar investment $V_t$ of the portfolio
(\ref{eq:contrarian}) and its unleveraged (Regulation T) portfolio return
$R_{pt}$ are given by:
\begin{eqnarray}
V_t &\equiv& \frac{1}{2}\sum_{i=1}^N |\omega_{it}|~~~,~~~R_{pt}\ \ \equiv\
\ \frac{\sum_{i=1}^N \omega_{it}R_{it}}{V_t}\ . \label{eq:return}
\end{eqnarray}
To construct leveraged portfolio returns $L_{pt}(\theta)$ using a
regulatory leverage
factor of $\theta\!:\!1$, we simply multiply (\ref{eq:return}) by $\theta/2$:%
\footnote{Note that Reg-T leverage is, in fact, considered 2:1 which is
exactly (\ref{eq:return}), hence $\theta\!:\!1$ leverage is equivalent to
a multiple of $\theta/2$.}
\begin{eqnarray}
L_{pt}(\theta) &\equiv& \frac{(\theta/2)\sum_{i=1}^N
\omega_{it}R_{it}}{V_t}\ .
\label{eq:leverage}
\end{eqnarray}

Because of the linear nature of the strategy, and Assumption (A1), the
strategy's statistical properties are particularly easy to derive.  For
example, Lo and MacKinlay (1990a) show that the strategy's profit-and-loss
at date $t$ is given by:
\begin{equation}
\label{eq:over-prft} \pi_t(k)\ \ =\ \ \bomega'_{t}(k) \bR_{t}
\end{equation}
and re-arranging (\ref{eq:over-prft}) and taking expectations yields the
following:
\begin{equation}
\label{eq:over-expct} {\rm E} [\pi_t(k)]\ \  =\ \
\frac{\biota'\bGamma_k\biota}{N^2}\ -\ \frac{1}{N}{\rm trace}(\bGamma_k) \
-\ \frac{1}{N}\sum_{i=1}^N (\mu_i - \mu_m)^2
\end{equation}
which shows that the contrarian strategy's expected profits are an
explicit function of the means, variances, and autocovariances of returns.
See Lo and MacKinlay (1990, 1999) for further details of this strategy's
statistical properties and a detailed empirical analysis of its historical
returns.

This initial phase of a strategy's development process is characterized by
the temporary fiction that Level-2 uncertainty holds, and not
surprisingly, we are able to derive fairly explicit and analytically
tractable results for the strategy's performance.  By making further
assumptions on the return-generating process $\{\bR_t\}$, we can determine
the profitability of this strategy explicitly.  For example, suppose that
the return of the $i$-th security is given by the following simple linear
relation:
\begin{equation}
R_{it}\ \ =\ \ \mu_i\ +\ \beta_i\Lambda_{t-i}\ +\ \epsilon_{it}~~,~~
\beta_i > 0~~,~~i=1,\ldots,N
\label{eq:leadlag}
\end{equation}
where $\epsilon_{it}$ is Gaussian white noise and $\Lambda_{t-i}$ is a
factor common to all stocks, but with a lag that varies according to the
index of security $i$, and which is also Gaussian and IID.  This
return-generating process is a special case of assumption (A1), and we can
compute the autocovariance matrices $\bGamma_k$ explicitly:
\begin{subeqnarray}
\bGamma_1 &=&
\left(~\begin{array}{cccccc}
0&\beta_1\beta_2&0&0&\cdots&0 \\
0&0&\beta_2\beta_3&0&\cdots&0 \\
\vdots&\vdots&\vdots&\vdots&\ddots&\vdots \\
0&0&0&0&\cdots&\beta_{N-1}\beta_N \\
0&0&0&0&\cdots&0 \\
\end{array}\right)\sigma^2_\lambda \qquad \\ [3ex]
\bGamma_2 &=&
\left(~\begin{array}{cccccc}
0&0&\beta_1\beta_3&0&\cdots&0 \\
0&0&0&\beta_2\beta_4&\cdots&0 \\
\vdots&\vdots&\vdots&\vdots&\ddots&\vdots \\
0&0&0&0&\cdots&\beta_{N-3}\beta_{N-1} \\
0&0&0&0&\cdots&0 \\
0&0&0&0&\cdots&0 \\
\end{array}\right)\sigma^2_\lambda\qquad \\ [3ex]
\bGamma_3 &=& \ldots \nonumber
\end{subeqnarray}
Therefore, for $k\!=\!1$, the expected profit of the strategy
(\ref{eq:contrarian}) is given by:
\begin{equation}
{\rm E} [\pi_t(1)]\ \ =\ \ \frac{\sigma^2_\lambda}{N^2}
\sum_{i=1}^{N-1} \beta_i\beta_{i+1}\ -\ \frac{1}{N}\sum_{i=1}^N (\mu_i -
\mu_m)^2
\label{eq:profits}
\end{equation}
If the $\beta_i$'s are of the same sign, and the cross-sectional
variability of the $\mu_i$'s is not too great, (\ref{eq:profits}) shows
that the strategy will be profitable on average.  By calibrating the
parameters of (\ref{eq:profits}) to a specific set of values, we can
evaluate the portfolio weights $\{\omega_{it}(k)\}$ and average
profitability ${\rm E}[\pi_t(1)]$ of the strategy (\ref{eq:contrarian})
explicitly.

\subsection{StatArb at Level 3}
\label{ssec:statarb_level3}
Having analyzed the mean-reversion strategy from a theoretical
perspective, a natural next step is to apply it to historical stock
returns and simulate its performance.  However, even the most cursory
review of the data will show some obvious inconsistencies between
assumption (A1) and reality, including non-normality and non-stationarities.%
\footnote{\label{fn:nonnormal} There is substantial evidence that 
financial asset returns are not normally distributed, but characterized by 
skewness, leptokurtosis, time-varying volatility and other non-Gaussian 
and non-stationarity properties (see, for example, Cootner (1964), 
Mantegna and Stanley, 1994, and Lo and MacKinlay, 1999).} 
In fact, the complexities of financial data and interactions make it
virtually impossible to determine the precise probability laws generating
the data, hence the best-case scenario we can hope for is Level-3
uncertainty when we engage in empirical analysis.  Fortunately, assumption
(A1) can be relaxed considerably---albeit at the expense of some
notational simplicity---by using approximation theorems such as the Law of
Large Numbers and the Central Limit Theorem, and using asymptotic
statistical inference in place of finite-sample methods.%
\footnote{For example, Lo and MacKinlay (1990a) show that the qualitative
features of their  analysis does not change under the weaker assumptions
of weakly dependent heterogeneously distributed vectors $\bR_t$ when
expectations are replaced with corresponding probability limits of
suitably defined time-averages.  See White for (1984) for further
details.}

With these caveats in mind, consider the empirical results reported by
Khandani and Lo (2007), who apply this strategy to the daily returns of
all stocks in the University of Chicago's CRSP Database, and to stocks
within 10 market-capitalization deciles, from January 3, 1995 to August
31,
2007.%
\footnote{Specifically, they use only U.S.\ common stocks (CRSP share code
10 and 11), which eliminates REIT's, ADR's, and other types of securities,
and they drop stocks with share prices below \$5 and above \$2{,}000. To
reduce unnecessary turnover in their market-cap deciles, they form these
deciles only twice a year (the first trading days of January and July).
Since the CRSP data are available only through December 29, 2006, decile
memberships for 2007 were based on market capitalizations as of December
29, 2006.  For 2007, they constructed daily close-to-close returns for the
stocks in their CRSP universe as of December 29, 2006 using adjusted
closing prices from {\tt finance.yahoo.com}. They were unable to find
prices for 135 stocks in their CRSP universe, potentially due to ticker
symbol changes or mismatches between CRSP and Yahoo.  To avoid any
conflict, they also dropped 34 other securities that are mapped to more
than one CRSP PERMNO identifier as of December 29, 2006. The remaining
3{,}724 stocks were then placed in deciles and used for the analysis in
2007. Also, Yahoo's adjusted prices do not incorporate dividends, hence
their 2007 daily returns are price returns, not {\it total returns}. This
difference is unlikely to have much impact on their analysis.}
Table \ref{tbl:contrarian} reports the year-by-year average daily return, 
standard deviation, and annualized Sharpe ratios (the annualized ratio of 
a strategy's expected return minus a riskfree rate of return to its return 
standard deviation) of (\ref{eq:contrarian}) and the results are 
impressive. In the first year of their sample, 1995, the strategy produced 
an average daily return of 1.38\% per day, or approximately 345\% per year 
assuming a 250-day year! The results for the following years are 
progressively less spectacular, but even in 2007, the average daily return 
of 0.13\% translates into a still-remarkable annualized return of 33\%! 
Moreover, the Sharpe ratios (computed under the assumption of a 0\% 
riskfree rate) are absurdly high at 53.87 in 1995, and still extreme at 
2.79 in 2007.  In comparison, the S\&P 500 had an annualized Sharpe ratio 
of 0.83 based on monthly returns from January 1995 to December 2007, and 
Warren Buffett's Sharpe ratio during this same period was 0.75 based on 
the daily returns of Berkshire Hathaway stock. As Khandani and Lo (2008, 
2009) observe, such performance is clearly too good to be true.  The 
absurdly profitable results suggest that Level-3 uncertainty does not 
fully characterize the nature of this strategy's performance, and we will 
return to this issue in Section \ref{ssec:statarb_level4}. 

\begin{table}[htbp]
\begin{center}
\includegraphics[clip, trim= 0 200 0 0,scale=.8]%
{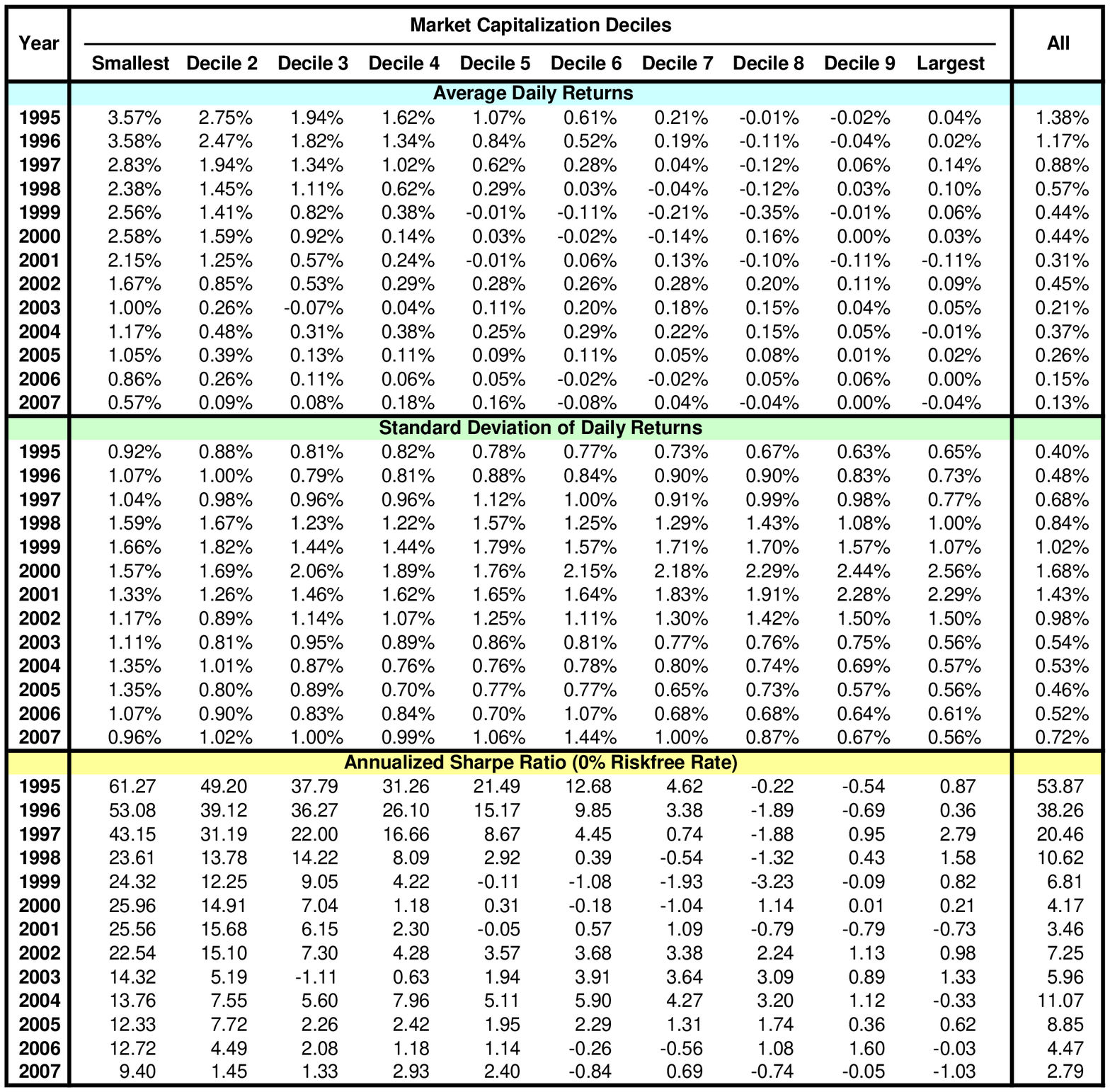}
\end{center}
\caption{Year-by-year average daily returns, standard deviations of daily
returns, and annualized Sharpe ratios ($\sqrt{250}\times (\textrm{average
daily return}/\textrm{standard deviation})$) of Lo and MacKinlay's (1990a)
contrarian trading strategy applied to all U.S.\ common stocks (CRSP share
codes 10 and 11) with share prices above \$5 and less than \$2{,}000, and
market-capitalization deciles, from January 3, 1995 to August 31, 2007.}
\label{tbl:contrarian}
\end{table}

Table \ref{tbl:contrarian} also confirms a pattern long recognized by
long/short equity managers---the relation between profitability and market
capitalization.  Smaller-cap stocks generally exhibit more significant
inefficiencies, hence the profitability of the contrarian strategy in the
smaller deciles is considerably higher than in the larger-cap portfolios.
For example, the average daily return of the strategy in the smallest
decile in 1995 is 3.57\% in contrast to 0.04\% for the largest decile. Of
course, smaller-cap stocks typically have much higher transactions costs
and price impact, hence they may not be as attractive as the data might
suggest.

\subsection{StatArb at Level 4}
\label{ssec:statarb_level4}
Of course, not even the most naive quant would take the results of Table 
\ref{tbl:contrarian} at face value.  There are many factors that must be 
weighed in evaluating these remarkable simulations, some easier to address 
than others.  For example, the numbers in Table \ref{tbl:contrarian} do 
not reflect the impact of transaction costs, which can be quite 
substantial in a strategy as active as (\ref{eq:contrarian}).  The high 
turnover and the large number of stocks involved underscores the 
importance that technology plays in strategies like (\ref{eq:contrarian}), 
and why funds that employ such strategies are predominantly quantitative. 
It is nearly impossible for human portfolio managers and traders to 
implement a strategy involving so many securities and trading so 
frequently without making substantial use of quantitative methods and 
technological tools such as automated trading platforms, electronic 
communications networks, and mathematical optimization algorithms. Indeed, 
part of the liquidity that such strategies seem to enjoy---the short 
holding periods, the rapid-fire implementation of trading signals, and the 
diversification of profitability across such a large number of 
instruments---is directly related to technological advances in trading, 
portfolio construction, and risk management.  Not surprisingly, many of 
the most successful funds in this discipline have been founded by computer 
scientists, mathematicians, and engineers, not by economists or 
fundamental stock-pickers. 

This reliance on technology and algorithms may give inexperienced quants
the mistaken impression that the mean-reversion strategy involves only
Level-3 uncertainty.  For example, trading costs are relatively easy to
incorporate into simulations since most costs are explicit and can
therefore be readily imputed.%
\footnote{Known costs include brokerage commissions, bid/offer spreads
(unless one is a broker-dealer, in which case the spread is a source of
revenue, not an explicit cost), exchange fees, ticket charges, implicit
net borrowing costs for any leverage employed by the strategy, and the
associated costs of the accounting, legal, systems, and telecommunications
infrastructure required to implement this strategy.}
However, certain implicit costs are harder to estimate, such as the impact 
of the strategy itself on market prices, i.e., the reaction of other 
market participants to the strategy's purchases and sales, and the 
ultimate effect on market prices and dynamics.  If the strategy is 
deployed with too much capital, financial markets tend to equilibrate by 
increasing the prices of the securities the strategy seeks to buy and 
decreasing the prices of the securities the strategy seeks to shortsell, 
thereby reducing the strategy's potential profits. If, in 1995, we had 
implemented the mean-reversion strategy with \$1 billion of capital and 
leveraged it 10:1, the very act of trading this strategy would likely have 
generated significant losses, due solely to the market's reaction to the 
strategy's large and frequent trades.  While such ```price impact'' is a 
fact of financial life, it is notoriously difficult to estimate accurately 
because it depends on many quantities that are beyond the portfolio 
manager's control and knowledge, including the identities of the manager's 
competitors on any given day, the strategies they are using, their 
business objectives and constraints, the likely reactions they might 
exhibit in response to the manager's trades, and how all of these factors 
may change from day to day as market conditions evolve.  This is but one 
example of Level-4 uncertainty. 

\begin{figure}[htbp]
\begin{center}
\includegraphics[clip, trim= 0 190 0 -175,scale=.55]%
{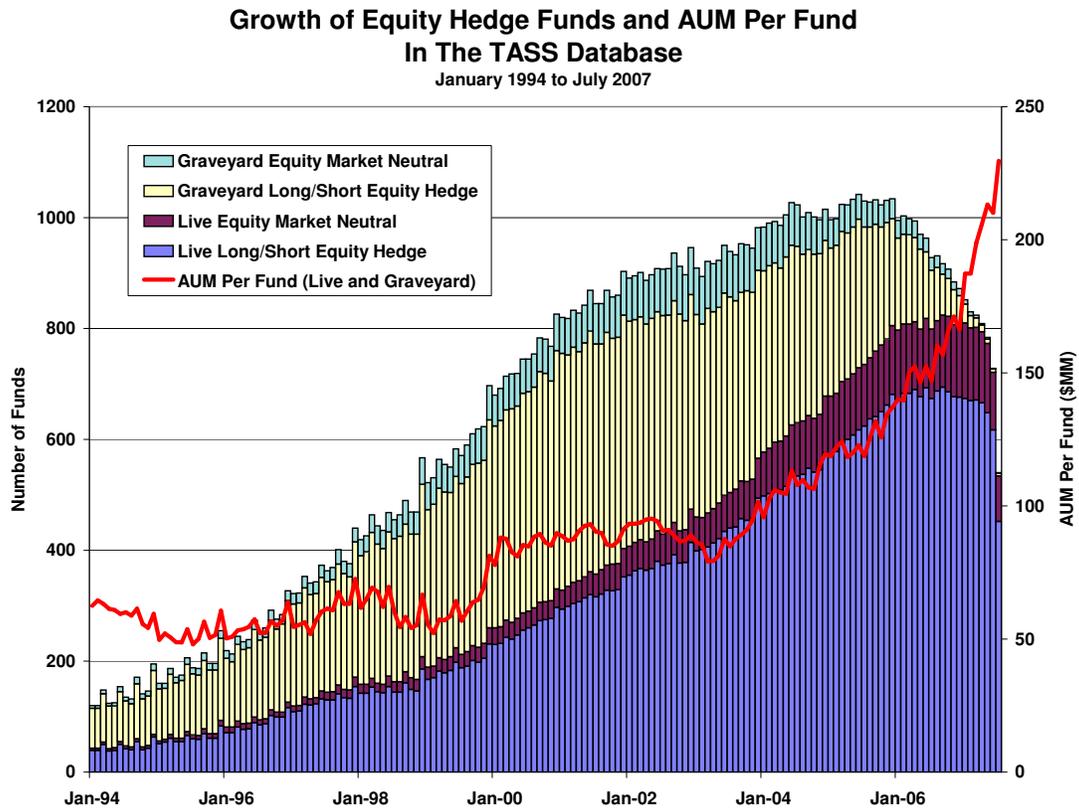}
\end{center}
\caption{Number of funds in the Long/Short Equity Hedge and Equity Market
Neutral categories of the TASS database, and average assets under
management per fund, from January 1994 to July 2007.} \label{fig:numfunds}
\end{figure}

A more troubling symptom of Level-4 uncertainty is the potential
non-stationarity of the strategy's returns implied by the near-monotonic
decline in the average daily return of the strategy from 1995 to 2007.
While certain types of non-stationarities such as time-varying
volatilities can be accommodated through more sophisticated econometric
techniques (see White, 1984), no statistical method can fully capture
wholesale changes in financial institutions and discrete structural shifts
in business conditions.  Of course, the strategy's declining profitability
over this period may simply be a statistical fluke, a random outcome due
to observational noise of an otherwise consistently profitable strategy.
Khandani and Lo (2007, 2008) propose a different hypothesis, one in which
the decline is symptomatic of the growing popularity of strategies such as
(\ref{eq:contrarian}), and a direct consequence of an enormous influx of
assets into such strategies during this period (see Figure
\ref{fig:numfunds}).  As more capital is deployed in similar strategies,
downward pressure is created on the strategies' returns.%
\footnote{Specifically, Khandani and Lo (2007, 2008) provide several
related causal factors for this trend: increased competition, changes in
market structure (e.g., decimalization), improvements in trading
technology and electronic connectivity, and the growth in assets devoted
to this type of strategy.}
This secular decline in profitability has significant implications for the
use of leverage, and the potential over-crowding that can occur among
trading strategies of this type.  We will return to this important issue
in the next section.

Statistical fluke or crowded trade---which of these two hypotheses is
closer to the truth?  This question is not merely of academic interest,
but is central to the business decision of whether or not to implement the
mean-reversion strategy (\ref{eq:contrarian}), how much capital to
allocate to it if implemented, and when to shut it down and whether and
when to restart it in the face of varying losses and gains.  A
particularly sobering episode of this simulated strategy's history is
August 2007---the main focus of Khandani and Lo (2007, 2008)---when the
strategy produced a cumulative
loss of approximately 25\% over a three-day period.%
\footnote{Khandani and Lo's (2007, 2008) simulation is quite involved, and
we refer readers to their papers for the full details.}
Given a daily standard deviation of 0.52\% for the strategy in 2006, the 
year prior to this event, and an $8\!:\!1$ leverage ratio assumed by 
Khandani and Lo (2007, 2008), a 25\% three-day loss represents a 
7-standard-deviation event, which occurs with probability $p\!=\!1.3\times 
10^{-12}$ assuming normality.  This probability implies that on average, 
such an event happens once every $6.4$ billion years, hence the last such 
event would be expected to have happened approximately $1.7$ billion years 
before the Earth was formed (see Lloyd, 2008). 

Events such as these seem to happen more than rarely, which is one of the 
reasons that some portfolio managers use ``stop-loss'' policies in which 
they reduce the size of the bets after experiencing a certain level of 
cumulative losses.  Such policies are often criticized as being ad hoc and 
``outside the model'', but implicit in this criticism is that the model is 
a complete description of the world, i.e., Level-3 uncertainty.  If there 
is some chance that the model does not capture every aspect of reality, 
and if that omitted aspect of reality can cause significant losses to a 
portfolio, then stop-loss policies may have their place in the quant 
manager's toolkit.%
\footnote{However, this observation does not imply that stop-loss policies 
should be used blindly.  During the Quant Meltdown of August 2007, 
Khandani and Lo (2007, 2008) observe that those managers who cut risk and 
unwound their positions during August 7--9 effectively locked in their 
losses and missed out on the sharp reversal of the strategy on August 10. 
The moral of this story is that the only way to avoid or reduce the impact 
of such extreme events is to develop a deeper understanding of the 
strategy's drivers, and in some cases, such understanding comes only after 
traumatic events like August 2007.} 

The point of these thought experiments is not to discredit mean-reversion
strategies or modern capitalism---many successful businesses and socially
valuable services have emerged from this milieu---but rather to provide a
concrete illustration of the impact of Level-4 uncertainty on financial
decisions.  Using a longer history of stock returns would do little to
shed light on the strategy's losses during August 2007; even an infinite
amount of data cannot reduce Level-4 uncertainty beyond a certain level
because of non-stationarities in financial markets and the economic
environment.  Yet financial decisions must be made even in the face of
Level-3 uncertainty, and managers, shareholders, investors, and regulators
must live with their consequences.

\section{Level-5 Uncertainty: Black Swan Song?}
\label{sec:swan}
Given the abstract nature of Level-5 uncertainty, and the fact that we did 
not consider this level explicitly in the examples of Sections 
\ref{sec:oscill} and \ref{sec:trading}, we expand on Level 5 in this 
section and provide some justification for its practical relevance.  In 
the context of the harmonic oscillator and statistical arbitrage, Level-5 
uncertainty cannot be addressed by estimating parameters in more 
sophisticated ways.  Instead, under this type of uncertainty, there is no 
plausible model or data-generating process that we can identify which 
systematically and accurately generates the prices we observe in financial 
markets.  Even if there are extended periods over which some version of 
our model is a useful and effective description of the data, we must be 
prepared for times when the limitations of the model are revealed, and 
human nature---in the form of fear, greed, and ``animal 
spirits''---appears in ways that few of us can anticipate. 

The Financial Crisis of 2007--2009 has generated renewed interest in this 
kind of uncertainty, which has come to be known in the popular media and 
within the finance community by a variety of evocative and often 
euphemistic names: ``tail events'', ``extreme events", or ``black swans'' 
(Taleb, 2007). These cultural icons refer to disasters that occur so 
infrequently that they are virtually impossible to analyze using standard 
statistical inference.  However, we find this perspective less than 
helpful because it suggests a state of hopeless ignorance in which we 
resign ourselves to being buffeted and battered by the unknowable. 

In this section, we explore this possibility by considering the entire 
range of uncertainty, initially from the rather unusual perspective of 
deterministic phenomena that are sometimes taken as random, and then in 
the context of financial markets and institutions.  We propose that the 
entire taxonomy of uncertainty must be applied at once to the different 
features of any single phenomenon, and these features can transition from 
one taxon to the next as we develop deeper understandings of the 
phenomenon's origins. 

\subsection{Eclipses and Coin Tosses}
\label{ssec:eclipses}
Consider an event that occurs with probability $p$ in any given year and
let $I_t$ be the 0/1 indicator variable of this event, hence:
\begin{equation}
I_t\ \ =\ \
\begin{cases}
1 & \mbox{with probability}~ p \\
0 & \mbox{with probability}~ 1\!-\!p
\end{cases}
\ .
\end{equation}
Let $\hat p$ denote the standard statistical estimator for $p$ based on
the historical relative frequency of the event in a sample of $T$
observations:
\begin{equation}
\hat p\ \ =\ \ \frac{1}{T}\sum_{t=1}^T I_t\ .
\end{equation}
The accuracy of $\hat p$ can be measured by computing its standard error,
which is given by:
\begin{eqnarray}
\SE[\hat p] \ \ =\ \ \sqrt{\Var[\hat p]}\ \ =\ \ \sqrt{\frac{1}{T} \Var[I_t]}
\ \ =\ \ \sqrt{\frac{p(1-p)}{T}}\ .
\label{eq:stderr}
\end{eqnarray}
Now consider a genuinely rare event, one the occurs once every 50 years,
hence $p\!=\!0.02$.  According to Table \ref{tbl:stderr}, even with 100
years of data, the standard error of the estimator $\hat p$ is 1.4\%, the
same order of magnitude as the estimator itself.  And this result has been
derived under ideal conditions; any serial correlation,
heteroskedasticity, measurement error, or nonstationarities in $\{I_t\}$
will only act to increase the standard errors of $\hat p$.

\begin{table}[htbp]
\centering
\includegraphics[clip, trim=0 625 0 0, scale=.8]%
{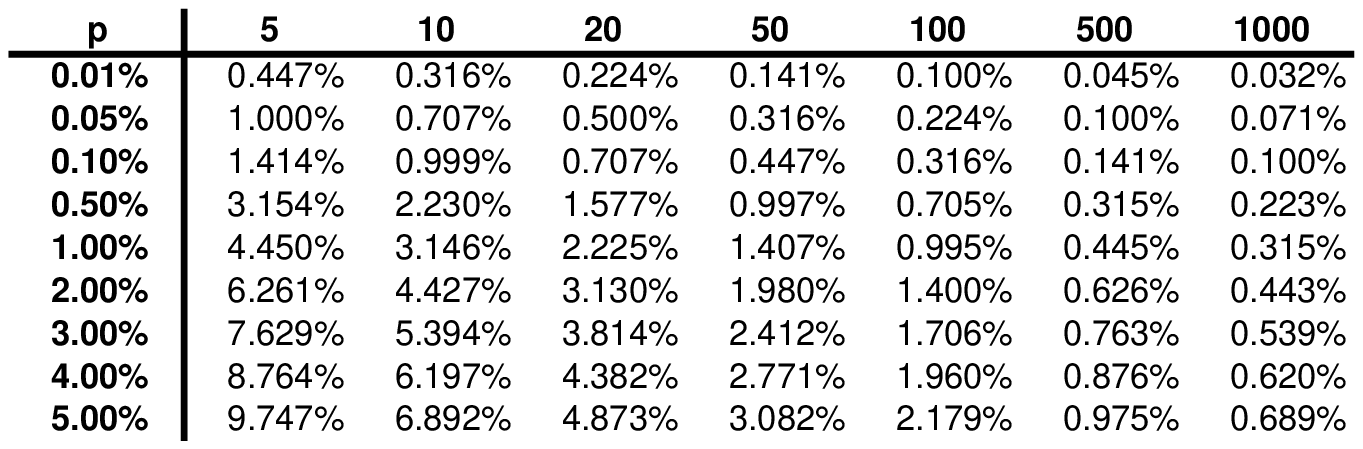}%
\caption{Standard errors of event probability estimator $\hat p$ as a
function of the actual probability $p$ and sample size $T$, assuming
independently and identically distributed observations.}
\label{tbl:stderr}
\end{table}

This simple calculation seems to suggest that rare events are difficult to
analyze by virtue of their rarity.  However, such a conclusion is a
reflection of the limitation of a pure statistical approach to analyzing
rare events, not a statement about rare events and irreducible
uncertainty.  For example, certain types of total solar eclipses (central,
with one limit) are exceedingly rare---only 49 will occur during the
10{,}000-year period from 4{,}000 BCE to 6{,}000 CE---yet they can be
predicted with extraordinary accuracy.%
\footnote{Eclipse predictions by Fred Espenak and Jean Meeus (NASA's GSFC)
are gratefully acknowledged.  See {\tt
http://eclipse.gsfc.nasa.gov/SEcatmax/SEcatmax.html} for further details.}
The language of probability and statistics is so well-developed and
ingrained in the scientific method that we often forget the fact that
probabilistic mechanisms are, in fact, proxies for deterministic phenomena
that are too complex to be modeled in any other fashion.%
\footnote{In a simple mathematical model of atmosphere convection, Lorenz
(1963) was perhaps the first to construct an example of a deterministic
process that exhibited complex behavior, i.e., sensitive dependence to
initial conditions.  Since then, the field of nonlinear dynamical systems
and ``chaos theory'' has identified numerous natural phenomena that are
chaotic (see, for example, Strogatz, 1994).}

\begin{figure}[htbp]
\centering
\includegraphics[clip, trim=0 260 0 0, scale=.8]%
{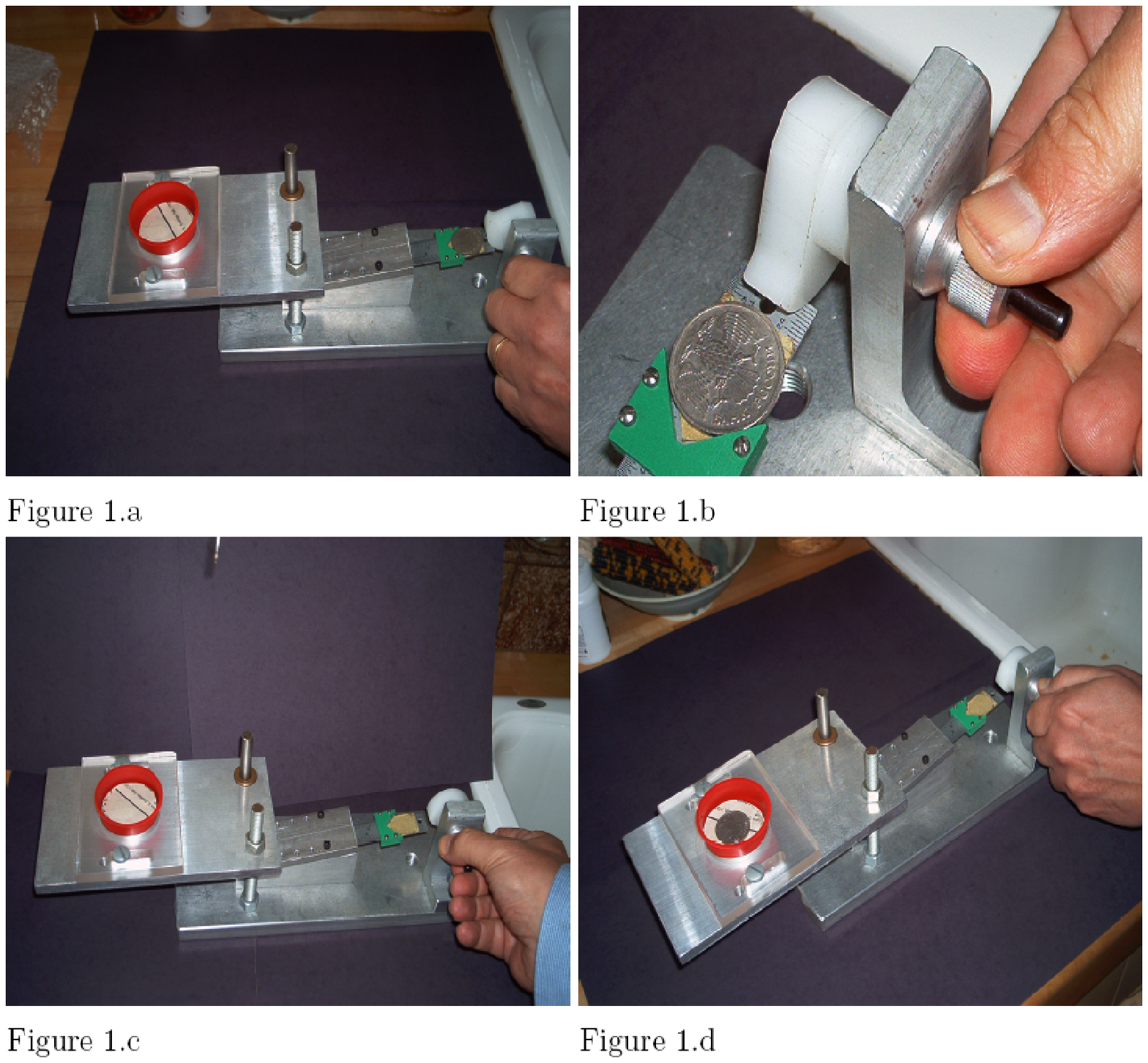}%
\caption{ Coin-tossing machine of Diaconis, Holmes, and Montgomery (2007, Figure
1): ``The coin is placed on a spring, the spring released by a ratchet,
the coin flips up doing a natural spin and lands in the cup. With careful
adjustment, the coin started heads up always lands heads up---one hundred
percent of the time. We conclude that coin-tossing is `physics' not
`random'''.}
\label{fig:heads}
\end{figure}

Perhaps the clearest illustration of this probabilistic approach to
deterministic complexity is the simple coin toss.  The 50/50 chances of
``heads or tails'' is used every day in countless situations requiring
randomization, yet it has been demonstrated conclusively that the outcome
of a coin toss is, in fact, deterministic, a product of physics not
probability (see Figure \ref{fig:heads} from Diaconis, Holmes, and
Montgomery, 2007).  Why, then, do we persist in treating the coin toss as
random?  The reason is that in all but the most sophisticated and
controlled applications, we have no idea what the initial conditions are,
nor does the typical coin-tosser understand enough of the physics of the
toss to make use of such initial conditions even if known.  Without the
proper data and knowledge, the outcome is effectively random.  But with
sufficient data and knowledge, the outcome is certain.

Within this simple example, we can observe the full range of uncertainty
from Level 5 to Level 1 just by varying the information available to the
observer.  Uncertainty is often in the eyes of the beholder, but we can
transition from one level of uncertainty to another as we deepen our
understanding of a given phenomenon.

\subsection{Uncertainty and Econometrics}
\label{ssec:econometrics}
An example of irreducible uncertainty more closely related to financial 
economics is provided by the econometric notion of identification.  If 
$\btheta$ is a vector of unknown parameters that characterize the 
probability law $\P(\bfX;\btheta)$ of a particular economic model with a 
vector of state variables $\bfX$, then the parameters are said to be 
``identified'' if and only if $\btheta_1\neq\btheta_2$ implies that 
$\P(\bfX;\btheta_1)\neq\P(\bfX;\btheta_2)$.  If this condition does not 
hold, then there is no way to distinguish between the two models 
characterized by $\btheta_1$ and $\btheta_2$ since they yield the same 
probability laws for the observable manifestation of the model, $\bfX$. 

A well-known application of this simple but powerful concept of 
identification is the econometric estimation of simultaneous linear 
equations (see Theil, 1971).  Consider the law of supply and demand which 
is taught in every introductory economics course: the price $P$ and 
quantity $Q$ that clear a market are determined by the intersection of the 
market's supply and demand curves.  Let the (linear) demand curve at time 
$t$ be given by: 
\begin{equation}
Q^d_t\ \ =\ \ d_0\ +\ d_1P_t\ +\ d_2Y_t\ +\ \epsilon^d_t
\label{eq:demand}
\end{equation}
where $P_t$ is the purchase/selling price, $Y_t$ is household income and
$\epsilon^d_t$ is a random demand shock, and let the (linear) supply curve
at time $t$ be given by:
\begin{equation}
Q^s_t\ \ =\ \ s_0\ +\ s_1P_t\ +\ s_2C_t\ +\ \epsilon^s_t
\end{equation}
where $C_t$ is production costs and $\epsilon^s_t$ is a random supply
shock. An economic equilibrium or market-clearing occurs when supply
equals demand, hence:
\begin{equation}
Q^s_t\ \ =\ \ Q^d_t\ \ =\ \ Q_t
\label{eq:equilibrium}
\end{equation}
where $Q_t$ is the equilibrium quantity produced and consumed at time $t$.

\begin{figure}[htbp]
\centering
\includegraphics[clip, trim=0 370 0 -140, scale=.6]%
{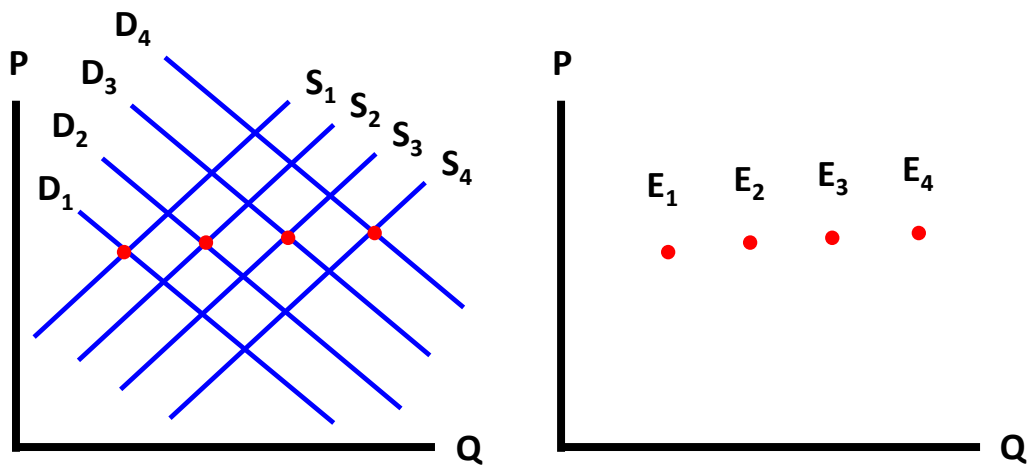}%
\caption{Illustration of the identification problem in the estimation of supply and
demand curves (left).  If only the intersections $E_i$ of the two curves
are observed (right), i.e., equilibrium prices and quantities, it is
impossible to determine the individual supply and demand curves generating
those equilibria, even with an infinite amount of data.}
\label{fig:supdem}
\end{figure}

Suppose that the variable to be forecasted is the quantity demanded
$Q^d_t$ as given by (\ref{eq:demand}),%
\footnote{This is not as frivolous an example as it may seem.  In
particular, a significant portion of consumer marketing research is
devoted to estimating demand curves.  Public policies such as gasoline,
alcohol, and cigarette taxes, energy tax credits, and investment
incentives also often hinge on price-elasticity-of-demand estimates.
Finally, economic analysis of demand is frequently the basis of damage
awards in a growing number of legal disputes involving securities fraud,
employment discrimination, and anti-trust violations.}
and we are given an infinite history of past prices $\{P_t\}$ and
quantities $\{Q_t\}$ observed in the marketplace, but no other
information.  It is easy to show that it is impossible to infer either the
demand or supply curves from this history, and forecasts of $Q^d_t$ will
be hopelessly unreliable because every single data point we observe is the
intersection of the two curves, hence there is no way to differentiate
between the two (see Figure \ref{fig:supdem}). In econometric terminology,
this system is not ``identified''.

Under the stated assumptions, achieving identification is quite simple.
Given sufficient historical data on household income $\{Y_t\}$ and
production costs $\{C_t\}$, a variant of linear regression---two-stage
least squares or instrumental variables---can accurately estimate both
supply and demand curves (Theil, 1971).  These additional variables allow
us to identify the two curves because they affect one curve but not the
other.  In particular, demand fluctuations due to variations in household
income can identify the supply curve, and supply fluctuations due to
variations in production costs can identify the demand curve. The
framework of supply and demand, and the additional data on household
income and production costs, allows us to move from Level-5 to Level-4
uncertainty.

As with the coin toss of Section \ref{ssec:eclipses}, this example 
contains the full range of our taxonomy of uncertainty, and illustrates 
the limitations of the black swan metaphor. Rather than simply admitting 
defeat and ascribing a particular outcome to a ``tail event'', we propose 
a more productive path.  By developing a deeper understanding of the 
underlying phenomena generating rare events, we are able to progress from 
one level of uncertainty to the next. 

\subsection{StatArb Revisited}
While the Quant Meltdown of August 2007 provides a compelling illustration 
of Level-4 uncertainty in a financial context, the analysis of Khandani 
and Lo (2007, 2008) offers some hope that additional quantitative 
investigation can change the proportion of irreducible uncertainty 
affecting mean-reversion strategies.  But what happened to these 
strategies from July to October 2008 may be more appropriately categorized 
as Level-5 uncertainty: in the wake of Bear Stearns' collapse and 
increasingly negative prospects for the financial industry, the U.S.\ 
Securities and Exchange Commission (SEC) temporarily prohibited the 
shorting of certain companies in the financial services sector, which 
eventually covered 799 securities (see SEC Release No.\ 34--58592). This 
unanticipated reaction by the government is an example of irreducible 
uncertainty that cannot be modeled quantitatively, yet has substantial 
impact on the risks and rewards of quantitative strategies like 
(\ref{eq:contrarian}). Even the most sophisticated simulation cannot 
incorporate every relevant aspect of market conditions as they evolve 
through time.  A truly thoughtful quant will understand this basic insight 
viscerally---having experienced losses first-hand thanks to Level-4 and 
Level-5 uncertainty---and will approach strategy development with humility 
and skepticism. 

The presence of Level-5 uncertainty should not, however, imply 
capitulation by the quant.  On the contrary, the need for analysis is even 
more urgent in these cases, but not necessarily of the purely quantitative 
type. Human intelligence is considerably broader than deductive logic, and 
other forms of reasoning can be effective when mathematics is not.  For 
example, in the case of the mean-reversion strategy (\ref{eq:contrarian}), 
it may be useful to observe that one possible source of its profits is the 
fact that the strategy provides liquidity to the marketplace.  By 
definition, losers are stocks that have underperformed relative to some 
market average, implying a supply/demand imbalance, i.e., an excess supply 
that caused the prices of those securities to drop, and vice-versa for 
winners. By buying losers and selling winners, mean-reversion strategies 
are adding to the demand for losers and increasing the supply of winners, 
thereby stabilizing supply/demand imbalances.  Traditionally, designated 
marketmakers such as the NYSE/AMEX specialists and NASDAQ dealers have 
played this role, for which they have been compensated through the 
bid/offer spread. But over the last decade, hedge funds and proprietary 
trading desks have begun to compete with traditional marketmakers, adding 
enormous amounts of liquidity to U.S.\ stock markets and earning 
attractive returns for themselves and their investors in the process. 

However, on occasion information affecting all stocks in the same
direction arises, temporarily replacing mean reversion with momentum as
the information is impounded into the prices of all securities.  In such
scenarios, liquidity providers will suffer large losses as this
information is processed by the market, and only after such information is
fully reflected in all prices will liquidity providers begin to profit
again.

This explanation by Khandani and Lo (2007, 2008) is plausible, but it is
not a purely quantitative hypothesis (although it certainly contains
quantitative elements).  Moreover, while certain components may be
empirically tested and either refuted or corroborated,%
\footnote{For example, in providing liquidity to the market,
mean-reversion strategies also have the effect of reducing market
volatility because they attenuate the movement of prices by selling stocks
for which there is excess demand and buying stocks for which there is
excess supply.  Therefore, an increasing amount of capital dedicated to
market-making strategies is one potential explanation for the secular
decline in U.S.\ equity-market volatility from 2003 to 2006. Once this
market-making capital is withdrawn from the marketplace, volatility should
pick up, as it did in the aftermath of the August 2007 Quant Meltdown.}
it may never be possible to determine the truth or falsity of their
conjectures.  This is Level-5 uncertainty.

\section{Applying the Taxonomy of Uncertainty}
\label{sec:apply}
While the taxonomy of uncertainty may be useful from an academic 
perspective, we believe that a clearer understanding of the types of 
uncertainty at play is critical to the proper \textit{management\/} of 
risk and uncertainty for all 
stakeholders in the financial system.%
\footnote{Although one of the central ideas of this paper is to
distinguish between risk and the various levels of uncertainty, we do not
expect standard terminology such as ``risk management''to reflect this
distinction, becoming ``risk and uncertainty management'' instead.
Accordingly, at the risk of creating some uncertainty, we shall conform
with standard terminology in this section.}
In this section, we describe some of these practical implications in more
detail.

In Section \ref{ssec:believe}, we consider the question asked by most 
investors of their quant portfolio managers: do you really believe your 
model?  The taxonomy of uncertainty suggests that the answer is not a 
simple yes or no.  This more nuanced perspective also applies to 
traditional methods of risk management, leading to the distinction between 
risk models and model risk discussed in Section \ref{ssec:riskmodelrisk}, 
the impact of misaligned incentives considered in Section 
\ref{ssec:moralhazard}, and the potential hazards of mismatched timescales 
in Section \ref{ssec:timescales}. We illustrate the practical relevance of 
all of these considerations in Section \ref{ssec:checklist} by proposing 
an uncertainty ``checklist'' that we apply to the quantitative strategy of 
Section \ref{sec:trading}. 

\subsection{Do You Really Believe Your Models??}
\label{ssec:believe}
Quantitative investment managers are often asked whether they ever 
over-ride their models on a discretionary basis.  Prospective investors 
are often more direct: ``Do you really believe your models??''.  One of 
the key ideas in applying the taxonomy of uncertainty is the fact that a 
given model typically exhibits multiple levels of uncertainty, hence an 
important ingredient in the successful implementations of any model is 
recognizing its domain(s) and, in particular, the boundaries of its 
validity.  Newtonian mechanics---while a remarkably useful description of 
a wide range of physical phenomena---is inadequate for many systems that 
require more sophisticated and encompassing theories, such as special 
relativity or quantum mechanics.  These exceptions do not cause us to 
abandon Newtonian mechanics, which forms the basis for every introductory 
physics course.  However, we acknowledge that developing new theories to 
extend the existing framework is a difficult, time-consuming, and 
worthwhile enterprise. 

In fact, it is possible to ``believe'' a model at one level of the
hierarchy but not at another.  Consider, for example, a forecasting model
for asset returns---often called an ``alpha model'' (such as the
mean-reversion strategy of Section \ref{sec:trading})---that seems
genuinely effective in generating positive profits on average after
deducting all transactions costs and adjusting for risk.%
\footnote{Risk adjustments are not without controversy, especially among
non-economists in the financial industry.  The basic idea is that positive
expected profits can be due to superior information and skill (alpha), or
due to particular risk exposures for which the positive expected profits
are the commensurate compensation (beta).  This distinction---first
espoused by Sharpe (1964) and Lintner (1965) and embodied in the ``Capital
Asset Pricing Model'' or CAPM---is important because it shows that
strategies with positive expected profits are not always based on
proprietary knowledge, but can sometimes be the result of risk premia.
However, there are now many versions of the CAPM, and hence many possible
ways to adjust for risk, and there is no consensus as to which one is
best.  Moreover, all of these adjustments require the assumption of
general equilibrium (the equality of supply and demand across all markets
at all times), which can also be challenged on many levels.}
Specifically, suppose that the historical or ``in-sample'' simulations of
this strategy generates a profit of 2\% per day with 55\% probability and
$-2\%$ per day with 45\% probability, and let these estimates be highly
statistically significant.  Despite the fact that this strategy
experiences losses 45\% of the time, over the course of a typical year,
its expected compound return is 65\%!  Moreover, suppose that in testing
this strategy on an ``out-of-sample'' basis using real capital and live
trading, the statistical properties of the strategy's realized returns are
similar to those of the backtest, after accounting for the inevitable
effects of backtest bias (see, for example, Lo and MacKinlay, 1990b).  In
this situation, it is understandable for the strategy's author to reply
``Yes, I do believe in my model''.

However, this response is likely based on the perspectives of Level-2 and 
Level-3 uncertainty, which may be the only perspectives of an 
inexperienced quant.  But the experienced quant also understands that 
statistical changes in regime, estimation errors, erroneous input data, 
institutional rigidities such as legal and regulatory constraints, and 
unanticipated market shocks are possible, i.e., Level-4 and Level-5 
uncertainty. In such circumstances, any given model may not only fail---it 
may fail spectacularly, as fixed-income arbitrage strategies did in August 
1998, as internet companies did in 2001--2002, as statistical arbitrage 
strategies did in August 2007, and as mortgage-backed securities did in 
2007--2009. 

There are two responses to the recognition that, in the face of Level-4 or 
Level-5 uncertainty, the model is outside of its domain of validity.  The 
first is to develop a deeper understanding of what is going on and to 
build a better model, which may be very challenging but well worth the 
effort if possible (see, for example, Lo and MacKinlay's, 1990a, analysis 
of the role of lead/lag effects in the profitability of Section 
\ref{sec:trading}'s mean-reversion strategy).  There are no simple 
panaceas; as with any complex system, careful analysis and creativity are 
required to understand the potential failures of a given model and the 
implications for a portfolio's profits and losses.  By developing a more 
complete understanding of the model's performance in a broader set of 
environments, we bring aspects of the model from Level-5 uncertainty to 
Level-4, from Level-4 to Level-3, and so on. 

The other response is to admit ignorance and protect the portfolio by 
limiting the damage that the model could potentially do.  This brings us 
to the subject of risk management, which lies at the heart of investment 
management. 

\subsection{Risk Models vs.\ Model Risk}
\label{ssec:riskmodelrisk}
Risk management always generates considerable support in concept, but its 
realities often differ from its ideals.  These discrepancies are largely 
attributable to the different levels of uncertainty.  To see why, let us 
return to the example in the previous section of the alpha model that 
yields $2\%$ each day with 55\% probability and $-2\%$ with 45\% 
probability.  Although its compound annual expected return is 65\%, this 
strategy's annualized return standard deviation is a whopping 53\% (in 
comparison, the annual standard deviation of the S\&P 500 is currently 
around 25\%). Such a high level of risk implies an 11\% chance that this 
highly profitable strategy's annual return is negative in any given year 
assuming Gaussian returns (which may not be appropriate; see footnote 
\ref{fn:nonnormal}), and a 30\% chance that over a 10-year period, at 
least two years will show negative returns.  Suppose in its first year of 
trading, the strategy experiences a $-50\%$ return---what is the 
appropriate response? 

From the perspective of Level-2 and Level-3 uncertainty, there is nothing
awry with such an outcome.  With an annual standard deviation of 53\%, an
annual return of $-50\%$ is well within the bounds of normality, both
figuratively and statistically.  Therefore, a plausible response to this
situation is to continue with the strategy as before, and this may well be
the correct response.

However, the perspective of Level-4 and Level-5 uncertainty suggests the
possibility that the losses are due to factors that are not so innocuous
and random.  While a $-50\%$ loss is not rare for a strategy with a $53\%$
annual standard deviation, it cannot help but change, to some degree, a
rational individual's prior belief that the strategy has a positive
expected return. Moreover, even if a loss of $50\%$ is statistically
common for such a strategy, there are business implications that may
intercede, e.g., the possibility of being shut down by senior management
or investors who do not wish to risk losing all of their capital.

More generally, much of the recent debate regarding the role of
quantitative models in the current financial crisis involves criticizing
standard risk analytics such as Value-at-Risk measures, linear factor
models, principal components analysis, and other statistical methods that
rely on assumptions of linearity, stationarity, or multivariate normality,
i.e., Level-2 and Level-3 uncertainty.  Under such assumptions, these
techniques can be readily implemented by estimating a covariance matrix
from the data, and then calculating the probabilities of gain and loss
based on the estimated parameters of a multivariate normal distribution
that incorporate volatility and correlation information.  The natural
conclusion drawn by the popular press is that such a naive approach to the
complexities of the real world caused the financial crisis (see the
discussion in Section \ref{sec:quants} on the Gaussian copula formula);
quants failed spectacularly.  In our view, this is a disingenuous
caricature that oversimplifies the underlying causes of the
crisis---giving too much weight to quants and too little to senior
management---and misses an important gap that exists in many risk
management protocols today.

There are certainly some quantitative analysts who have enough experience 
to understand the business end of trading desks and portfolio management. 
Similarly, there are certainly some senior managers who have more than a 
passing familiarity with the models that are used to calculate the risk 
exposures of their positions.  But when, in a single organization, the 
specialization of these two perspectives becomes too great---with quants 
focusing on Level-1 certainty and Level-2 uncertainty, and senior managers 
focusing on Level-3 and Level-4 uncertainty---a potentially dangerous 
risk-management gap is created. This gap usually emerges gradually and 
without notice, perhaps through the departure of quantitatively 
sophisticated senior management, the rapid promotion of business managers 
without such training, or the rising influence of inexperienced quants who 
happened to have enjoyed a period of profitable trading.  The appropriate 
response to this gap is neither to discard quantitative models entirely, 
nor to defend them to the death, but rather to acknowledge their 
limitations and develop additional methods for addressing them. 

One approach is to develop a set of risk-management protocols that may lie
outside of the framework of the alpha model.  In contrast to the typical
linear factor models that use factors drawn from the alpha model to
estimate a strategy's risk exposures, volatility, Value-at-Risk
statistics, and other traditional risk analytics, we are referring to a
``meta-model'' that can account for ``alpha decay'' (the declining
profitability of a strategy) and ``model risk'' (the complete failure of a
model due to circumstances outside the model's set of assumptions).  In
other words, a complete risk management protocol must contain risk models
but should also account for model risk.  Specific methods for dealing with
model risk include stop-loss policies that incorporate the risk of ruin or
franchise risk; limits on capital commitments and positions (to prevent
``outlier'' signals from making unusually large or concentrated bets);
statistical regime-switching models that can capture changes in the
profitability of the strategy; and Bayesian decision rules to trade off
``Type I and Type II errors'' (making a bet when the signal is wrong vs.\
not making a bet when the signal is right).

These are crude but often effective responses to Level-5 uncertainty, 
whether they come in the form of counterparty failure, sovereign or 
corporate defaults, currency devaluations, accounting fraud, rapid shifts 
in regulatory requirements or capital controls, political and social 
upheaval, a change in business strategy or corporate control, or any other 
event that is difficult or impossible to anticipate.  Such policies are 
closer to business decisions than trading decisions---but, of course, one 
important objective of trading is to stay in business---and requires 
managers who understand both the business environment and the limitations 
of models and the consequences of their failure.  When senior management 
is missing one of these two perspectives, the consequences can be 
disastrous. 

\subsection{Incentives and Moral Hazard}
\label{ssec:moralhazard}
A more prosaic example of Level-4 uncertainty in managing the risks of
proprietary trading strategies such as (\ref{eq:contrarian}) is the role
of incentives.  The so-called ``principal-agent'' problem of delegated
portfolio management is well-known (see Stiglitz, 1987), but this
challenge is easier to state than to solve in a real-world context.
Consider the following thought experiment: you are given an investment
opportunity in which you stand to earn \$10 million with 99.9\%
probability, but with 0.1\% probability, you will be executed.  Is it
clear what the ``right'' decision is?  Now suppose it is your boss that is
to be executed instead of you; does that change your decision?  Similar
choices---perhaps with less extreme downsides---are presented everyday to
portfolio managers in the financial industry, and the incentive structures
of hedge funds, proprietary trading desks, and most non-financial
corporations have a non-trivial impact on the attendant risks those
financial institutions face.

A broader challenge regarding incentives involves the typical corporate
governance structure in which risk management falls under the aegis of the
CEO and CIO, both of whom are focused on maximizing corporate profits, and
where risk is viewed primarily as a constraint.  Moreover, most Chief Risk
Officers (CROs) are compensated in much the same way that other senior
management is: through stock options and annual bonuses that depend, in
part, on corporate profits.  One alternative is for a corporation to
appoint a Chief Risk Officer (CRO) who reports directly to the board of
directors and is solely responsible for managing the company's enterprise
risk exposures, and whose compensation depends not on corporate revenues
or earnings, but on corporate stability.  Any proposed material change in
a corporation's risk profile---as measured by several objective metrics
that are specified in advance by senior management and the board---will
require prior written authorization of the CRO, and the CRO can be
terminated for cause if a corporation's risk profile deviates from its
pre-specified risk mandate, as determined jointly on an annual basis by
senior management and the board.

Such a proposal does invite conflict and debate among senior management
and their directors, but this is precisely the point.  By having open
dialogue about the potential risks and rewards of new initiatives---which
would provide a natural forum for discussing Level-4 and Level-5
uncertainty, even among the most quantitatively sophisticated
models---senior management will have a fighting chance of avoiding the
cognitive traps that can lead to disaster.

\subsection{Timescales Matter}
\label{ssec:timescales}
Another practical consideration that must figure in any application of our
taxonomy of uncertainty is the natural timescale of the strategy and how
it may differ from the timescale of the various decisionmakers involved.
In some cases, a significant mismatch in timescales can lead to disaster.
For example, if a strategy is based on quarterly earnings information and
a portfolio of positions designed to exploit such information is updated
only once a quarter, it cannot adjust to sharp changes in volatility that
may occur within the quarter.  Accordingly, such a portfolio may contain
more risk than its manager or investors expect.  Similarly, if financial
regulations are updated only once every few years, it is easy to see how
such laws can easily be outpaced and outmaneuvered by more rapid financial
innovation during periods of rapid economic growth.

A simple and intuitive heuristic is to set the timescale of investment
decisions and risk management to be at least as fine as the finest time
interval in which information relevant to the strategy is generated. If,
for example, daily signals are employed in a strategy (such as the
strategy in Section \ref{sec:trading}), then daily risk analytics,
stop-loss policies, and position limits are necessary; monthly versions
would be nearly useless.  On the other hand, a monthly portfolio strategy
may still require daily rebalancings and analytics if intra-monthly
volatility spikes or large profit-and-loss swings affect the strategy's
operations or business viability.  In contrast, the case of legendary
investor Warren Buffett's portfolio---which is constructed under a
timescale of years over which its investments are expected to pay off---is
unlikely to require monthly rebalancings (although Buffett does have the
extra advantage of very deep pockets, which is perhaps the most effective
risk management tool of all).

Alternatively, when the timescale of the uncertainty matches the timescale
of investment decisions, the challenges are greatest, as the
regime-switching harmonic oscillator of Section \ref{ssec:osc3}
illustrates.  If the timescale of uncertainty is much greater than that of
the investment decision (e.g., the impact of climate change on the
profitability of statistical arbitrage strategies) or much less (e.g., the
impact of restricting high-frequency trading on Buffett's portfolio), such
uncertainty can usually be ignored as a first approximation.  But when the
timescale of the uncertainty is comparable to the timescale of the
investment decisions, extreme caution must be exercised in managing such
uncertainty.  In such circumstances, a proper understanding of the
taxonomy of uncertainty and how it applies to every aspect of an
investment strategy becomes critical.

\subsection{The Uncertainty Checklist}
\label{ssec:checklist}
To illustrate the practical relevance of the taxonomy of uncertainty, we
propose a simple ``uncertainty checklist'' that can be applied to any
business endeavor.  In addition to providing a compact summary of the
implications of the five levels of uncertainty, the checklist also serves
as a useful starting point for designing a comprehensive enterprise risk
management protocol that goes well beyond the standard framework of VaR,
stress tests, and scenario analysis.

The idea of an uncertainty checklist is straightforward: it is organized
as a table whose columns correspond to the five levels of uncertainty of
Section \ref{sec:taxonomy}, and whose rows correspond to all the business
components of the activity under consideration.  Each entry consists of
all aspects of that business component falling into the particular level
of uncertainty, and ideally, the individuals and policies responsible for
addressing their proper execution and potential failings.

For concreteness, consider the quantitative trading strategy described in
Section \ref{sec:trading}.  The components of such a strategy might
include:
\begin{itemize}
\itemsep -0.0675in
\item theoretical framework (alpha models and forecasts);
\item empirical analysis (data and backtest platform);
\item portfolio construction (optimization engine);
\item trading and implementation (market frictions, telecommunications,
operations);
\item risk management (limits on positions, VaR, and losses);
\item legal and regulatory (trading restrictions, reporting requirements);
and
\item business issues (funding sources, investor relations, economic
conditions).
\end{itemize}
Table \ref{tbl:checklist} contains an example of what an uncertainty 
checklist might include; the entries are by no means complete, but are 
meant to illustrate the type of issues to be considered at each level of 
uncertainty and for each component of the business.  This example is also 
meant to convey the multi-dimensional nature of truly comprehensive 
enterprise-risk-management protocols, and the different skills and systems 
needed to address the various cells of this two-dimensional checklist.  In 
particular, issues involving the upper left portion of the checklist 
require strong quantitative skills, whereas issues involving the lower 
right portion require deep business experience and intuition.  But this 
table highlights the critical fact that the business involves all of these 
cells, which any risk management protocol must integrate to be effective. 

\begin{table}[htbp]
\begin{center}
\includegraphics[clip, trim= 0 165 0 0,scale=.8]%
{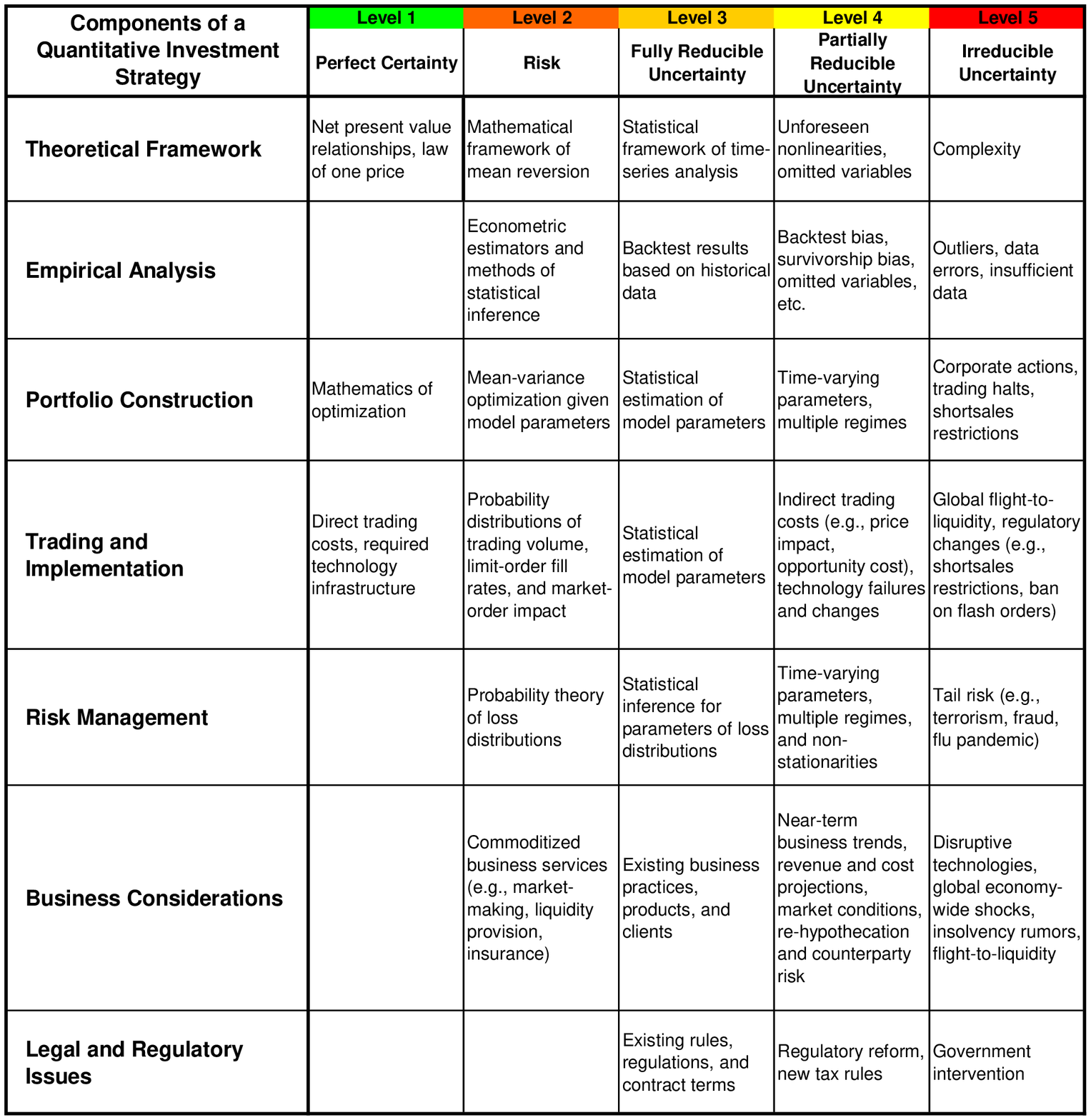}
\end{center}
\caption{The uncertainty checklist as applied to a quantitative equity
trading strategy.} \label{tbl:checklist}
\end{table}

The many facets of uncertainty can even affect an activity as mundane as
cash management.  At the start of a new trading strategy, the amount of
capital available for deployment is usually known with certainty (Level
1). However, over time, capital inflows and outflows can occur, injecting
some randomness---typically small in relation to the total amount of
capital deployed---into the day-to-day sizing of the strategy (Levels 2
and 3).  On occasion, extreme positive or negative performance can
generate large capital flows over short time intervals, creating
significant trading and implementation challenges with outcomes that
cannot be completely quantified (Level 4).  And because certain
contractual agreements allow counterparties the flexibility to re-assign
obligations (swap novation) or to re-use collateral in unspecified ways
with unspecified third parties (re-hypothecation), it is sometimes
impossible to determine even the qualitative nature of one's risk
exposures (Level 5).  While cash management may seem trivial and
pedestrian from the portfolio manager's perspective, those who had
significant cash holdings at Lehman Brothers in 2008 may now have a deeper
appreciation of its subtleties and hidden dangers.

The purpose of the uncertainty checklist is not simply to catalog 
potential sources of disaster, but rather to develop a systematic approach 
to risk management that acknowledges the presence of distinct types of 
uncertainty, each requiring a different set of skills and systems to 
manage properly.  For many of the cells in Table \ref{tbl:checklist}, 
there may be little to do other than to admit ignorance, but even this can 
be valuable from the perspective of corporate strategy.  The very process 
of completing the uncertainty checklist may generate important questions, 
concerns, and insights that might otherwise not have surfaced during 
business as usual. 

\section{Quants and the Current Financial Crisis}
\label{sec:quants}
No discussion of the future of finance would be complete without
considering the role that quantitative models may have played in creating
the current financial crisis. There is, of course, little doubt that
securitization, exotic mortgages, credit default swaps, credit ratings for
these new instruments, and other financial innovations that depended on
quantitative models were complicit. But so too were homeowners, mortgage
brokers, bankers, investors, boards of directors, regulators, government
sponsored enterprises, and politicians.  There is plenty of blame to go
around, and the process of apportioning responsibility has only just
begun.  But blaming quantitative models for the crisis seems particularly
perverse, and akin to blaming arithmetic and the real number system for
accounting fraud.

The typical scientist's reaction to failure is quite the opposite of 
vilification or capitulation.  Faced with the complete failure of 
classical physics to explain the spectrum of electromagnetic radiation in 
thermal equilibrium, or the energies of electrons ejected from a metal by 
a beam of light, physicists in the opening decades of the twentieth 
century responded not by rejecting quantitative analysis, but by studying 
these failures intensely and redoubling their efforts to build better 
models.  This process led to the remarkable insights of Planck and 
Einstein and, ultimately, to an entirely new model of Nature: quantum 
mechanics. In fact, in every science, knowledge advances through the 
painstaking process of successive approximation, typically with one 
generation of scholars testing, refining, or overturning the models of the 
previous generation. As Newton put it, ``If I have seen further it is only 
by standing on the shoulders of giants''.  Samuelson, paraphrasing 
Planck's, was somewhat darker: ``Progress in the sciences occurs funeral 
by funeral''. 

Among the multitude of advantages that physicists have over financial 
economists is one that is rarely noted: the practice of physics is largely 
left to physicists.  When physics experiences a crisis, physicists are 
generally allowed to sort through the issues by themselves, without the 
distraction of outside interference.  When a financial crisis occurs, it 
seems that everyone becomes a financial expert overnight, with 
surprisingly strong opinions on what caused the crisis and how to fix it. 
While financial economists may prefer to conduct more thorough analyses in 
the wake of market dislocation, the rush to judgment and action is 
virtually impossible to forestall as politicians, regulators, corporate 
executives, investors, and the media all react in predictably human 
fashion.  Imagine how much more challenging it would have been to fix the 
Large Hadron Collider after its September 19, 2008 short circuit if, after 
its breakdown, Congress held hearings in which various 
constituents---including religious leaders, residents of neighboring 
towns, and unions involved in the accelerator's construction---were asked 
to testify about what went wrong and how best to deal with its failure. 
Imagine further that after several months of such hearings, 
politicians---few of whom are physicists---start to draft legislation to 
change the way particle accelerators are to be built, managed, and 
staffed, and compensation limits are imposed on the most senior research 
scientists associated with the facility.

Although there are, of course, many differences between the financial 
system and a particle accelerator, the point of our analogy is to 
underscore the dangers of reacting too quickly to crisis in the case of 
highly complex systems.  Today's financial system is considerably more 
complex than ever before, requiring correspondingly greater concerted 
effort and expertise to overhaul.  By comparison, the Large Hadron 
Collider---as complex as it is---is a \emph{much\/} simpler system.  
Although it has been designed to probe new concepts beyond the current 
Standard Model of elementary particle physics, the accelerator, detectors, 
and computers that constitute the Collider are governed by physical laws 
that are either 
perfectly certain or fully reducibly uncertain.%
\footnote{When we claim that these laws are perfectly certain or have 
fully reducible uncertainty, we are referring to the forms of the laws and 
the statistical nature of the predictions they make.  Quantum mechanics is 
intrinsically probabilistic in predicting the outcomes of individual 
experiments, but the form of the Schr\"odinger equation is not 
probabilistic, and repeated experiments lead to well-defined statistical 
outcomes.  More formally, the functioning of superconducting magnets and 
calorimeters can be accurately represented as having a much smaller degree 
of partially-reducible uncertainty than statistical (fully-reducible) 
uncertainty, which is the essential point of this simple analogy.} 
Yet a collaboration of over 10{,}000 scientists was required to design and
build it, and the September 19, 2008 short circuit took over a year to
troubleshoot and repair, having come back on line on November 20, 2009.

To underscore the importance of the scientific method in analyzing the
current financial crisis, in this section we present three commonly held
beliefs about the crisis that, after more careful scrutiny, are
demonstrably untrue or unwarranted. Our purpose is not to criticize those
who hold such beliefs, but rather to remind ourselves that one of the
drawbacks of human creativity is the remarkable ease with which we draw
conclusions in the absence of facts.  This tendency is especially common
in emotionally charged situations, perhaps because the urgency of action
is that much greater (see de Becker, 1997). Scientific inquiry is a
potent, and perhaps the only, antidote.

\subsection{Did the SEC Allow Too Much Leverage?}

On August 8, 2008, the former director of the SEC's Division of Market 
Regulation (now the ``Division of Markets and Trading''), Lee Pickard, 
published an article in the {\it American Banker\/} with a bold claim: a 
rule change by the SEC in 2004 allowed broker-dealers to greatly
increase their leverage, contributing to the financial crisis.%
\footnote{We thank Jacob Goldfield for bringing this example to our 
attention.}
In particular, Mr. Pickard argued that before the rule change, ``the 
broker-dealer was limited in the amount of debt it could incur, to about 
12 times its net capital, though for various reason broker-dealers 
operated at significantly lower ratios$\,\ldots$ If, however, Bear Stearns 
and other large broker-dealers had been subject to the typical haircuts on 
their securities positions, an aggregate indebtedness restriction, and 
other provisions for determining required net capital under the 
traditional standards, they would not have been able to incur their high 
debt leverage without substantially increasing their capital base.''\ \ He 
was referring to a change in June 2004 to SEC Rule 15c3--1, the so-called 
``net capital rule'' by which the SEC imposes net capital requirements 
and, thereby, limits the leverage employed by broker-dealers. This serious 
allegation was picked up by a number of newspapers, including the {\it New 
York Times} on October 3, 2008 (Labaton, 2008): 
\begin{quote}
\baselineskip=14pt%
In loosening the capital rules, which are supposed to provide a buffer in
turbulent times, the agency also decided to rely on the firms' own
computer models for determining the riskiness of investments, essentially
outsourcing the job of monitoring risk to the banks themselves.

Over the following months and years, each of the firms would take
advantage of the looser rules. At Bear Stearns, the leverage ratio---a
measurement of how much the firm was borrowing compared to its total
assets---rose sharply, to 33 to 1. In other words, for every dollar in
equity, it had \$33 of debt. The ratios at the other firms also rose
significantly.
\end{quote}
These reports of sudden increases in leverage from 12-to-1 to 33-to-1 
seemed to be the ``smoking gun'' that many had been searching for in their 
attempts to determine the causes of the Financial Crisis of 2007--2009.  
If true, it implied an easy fix according to Pickard (2008): ``The SEC 
should reexamine its net capital rule and consider whether the traditional 
standards should be reapplied to all broker-dealers.'' 

While these ``facts'' seemed straightforward enough, it turns out that the 
2004 SEC amendment to Rule 15c3--1 did nothing to change the leverage 
restrictions of these financial institutions. In a speech given by the 
SEC's director of the Division of Markets and Trading on April 9, 2009 
(Sirri, 2009), Dr.\ Erik Sirri stated clearly and unequivocally that 
``First, and most importantly, the Commission did not undo any leverage 
restrictions in 2004''.%
\footnote{SEC Rule 15c3--1 is complex, and not simply a leverage test.  
The rule does contain a 15-to-1 leverage test with a 12-to-1 ``early 
warning'' obligation.  However, this component of the rule only limits 
unsecured debt, and did not apply to large broker-dealers, who were 
subject to net capital requirements based on amounts owed to them by their 
customers, i.e., a customer-receivable or ``aggregate debit  item'' test.  
This test requires a broker-dealer to maintain net capital equal to at 
least 2\% of those  receivables, which is how the five large investment 
banks had been able to achieve higher leverage ratios in the 1990's than 
after the 2004 rule change (see Figure \ref{fig:GAOlev} below).  
Similarly, their broker-dealer  subsidiaries (which were the entities 
subject to the net capital rule) had long achieved leverage ratios far in 
excess of 15-to-1.  The historical leverage ratios of the investment banks 
were readily available in their financial reports, and the facts regarding 
the true nature of the SEC net capital rule were also available in the 
public domain. We thank Bob Lockner for helping us with the intricacies of 
the SEC net capital rule.} 
He cites several documented and verifiable facts to support this 
surprising conclusion, and this correction was reiterated in a letter from 
Michael Macchiaroli, Associate Director of the Division of Markets and 
Trading to the General Accountability Office (GAO) on July 17, 2009, and 
reproduced in the GAO Report GAO--09--739 (2009, p.~117). 

What about the stunning 33-to-1 leverage ratio reported by the press?  
According to the GAO (Report GAO--09--739, 2009, p.~40): 
\begin{quote} 
In our prior work on Long-Term Capital Management (a hedge fund), we 
analyzed the assets-to-equity ratios of four of the five broker-dealer 
holding companies that later became CSEs and found that three had ratios 
equal to or greater than 28-to-1 at fiscal year-end 1998, which was higher 
than their ratios at fiscal year-end 2006 before the crisis began (see 
fig.\ 6).
\end{quote}
\noindent
In footnote 68 of that report, the GAO observes that its 1999 report 
GAO/GGD--00--3 (1999) on Long-Term Capital Management ``$\ldots$ did not 
present the assets-to-equity ratio for Bear Stearns, but its ratio also 
was above 28 to 1 in 1998''.  The GAO's graph of the historical leverage 
ratios for Goldman Sachs, Merrill Lynch, Lehman Brothers, and Morgan 
Stanley is reproduced in Figure \ref{fig:GAOlev}.  These leverage numbers 
were in the public domain and easily accessible through company annual 
reports and quarterly SEC filings.

\begin{figure}[htbp]
\centering
\includegraphics[clip, trim=0 200 0 -160, scale=.55]%
{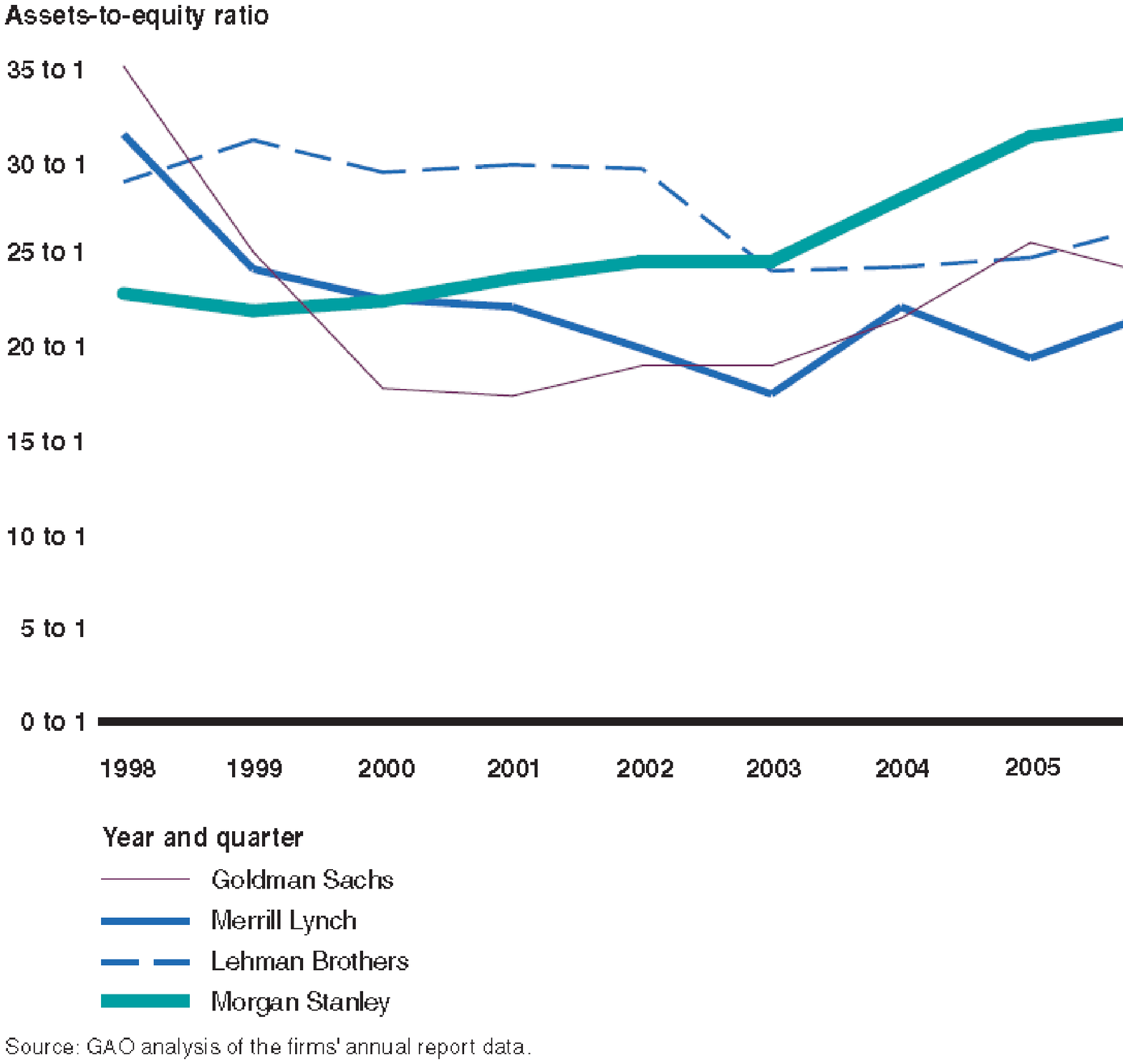}%
\caption{Ratio to total assets to equity for four broker-dealer holding
companies from
1998 to 2007.  Source: U.S.\ Government Accountability Office 
Report GAO--09--739 (2009, Figure 6).}
\label{fig:GAOlev}
\end{figure}
 
Now it must be acknowledged that the arcane minutiae of SEC net capital 
rules may not be common knowledge, even among professional economists, 
accountants, and regulators.  But two aspects of this particular incident 
are noteworthy: (1)~the misunderstanding seems to have originated with the 
allegation by Mr.~Pickard, who held the same position as Dr.~Sirri at the 
SEC from 1973--1977 and was involved in writing the original version of 
Rule 15c3--1; and (2)~the allegation was used by a number of prominent and 
highly accomplished economists, regulators, and policymakers to formulate 
policy recommendations without verifying the scientific merits of the 
claim that ``changes in 2004 net capital rules caused investment banks
to increase leverage''.%
\footnote{For example, former SEC chief economist Susan Woodward cited 
this ``fact'' in a presentation to the Allied Social Sciences Association 
Annual Meeting (the largest annual gathering of economists) on January 3, 
2009 (Woodward, 2009).  On December 5, 2008, Columbia University Law 
Professor John Coffee (2008) wrote in the {\it New York Law Journal} that 
after the rule change,``The result was predictable: all five of these 
major investment banks increased their debt-to-equity leverage ratios 
significantly in the period following their entry into the CSE program''. 
In a January 2009 {\it Vanity Fair\/} article, Nobel-prize-winning 
economist Joseph Stiglitz (2009) listed five key ``mistakes'' that led to 
the financial crisis and ``One was the decision in April 2004 by the 
Securities and Exchange Commission, at a meeting attended by virtually no 
one and largely overlooked at the time, to allow big investment banks to 
increase their debt-to-capital ratio (from 12:1 to 30:1, or higher) so 
that they could buy more mortgage-backed securities, inflating the housing 
bubble in the process''.  And on January 24, 2009, Alan Blinder, a 
professor of economics at Princeton University and a former vice-chairman 
of the U.S.\ Federal Reserve, published a piece in the ``Economic View'' 
column of the {\it New York Times\/} titled ``Six Errors on the Path to 
the Financial Crisis'' in which this rule change was Error Number 2 on his 
list (Blinder, 2009).}
If such sophisticated and informed individuals can be so easily misled on
a relatively simple and empirically verifiable issue, what does that imply
about less-informed stakeholders in the financial system?

This unfortunate misunderstanding underscores the need for careful 
deliberation and analysis, particularly during periods of extreme distress 
when the sense of urgency may cause us to draw inferences and conclusions 
too quickly and inaccurately.  We conjecture that the misunderstanding was 
generated by the apparent consistency between the extraordinary losses of 
Bear, Lehman, and Merrill and the misinterpretation of the 2004 SEC rule 
change---after all, it seems perfectly plausible that the apparent 
loosening of net capital rules in 2004 could have caused broker-dealers to 
increase their leverage. Even sophisticated individuals form mental models 
of reality that are not complete or entirely accurate, and when new 
information is encountered, our cognitive faculties are hardwired to 
question first those pieces that are at odds with our mental model.  When 
information confirms our preconceptions, we usually do not ask why. 

As of March 12, 2010, the {\it New York Times\/} has yet to print a 
correction of its original stories on the 2004 change to Rule 15c3--1, nor 
did the {\it Times\/} provide any coverage Dr.~Sirri's April 9, 2009 
speech.  Apparently, correcting mistaken views and factual errors is not 
always  news.  But it is good science. 

\subsection{If Formulas Could Kill}
At the epicenter of the current financial crisis are bonds backed by large
pools of geographically diversified residential mortgages and stratified
into ``tranches'' by credit quality.  Such securities are examples of
``asset-backed securities'' (ABSs) or, when the tranches of several ABSs
are pooled, ``collateralized debt obligations'' (ABS-CDOs).  ABS-CDOs have
been used to securitize a wide variety of assets in addition to mortgages,
such as credit card receivables, student loans and home equity loans.  The
mortgage-backed bonds reflected the aggregate risk/return characteristics
of the cashflows of the underlying mortgages on which the bonds' cashflows
were based.  By pooling mortgages from various parts of the country,
issuers of ABS-CDOs were diversifying their homeowner default risk in
precisely the same way that an equity index fund diversifies the
idiosyncratic risks associated with the fortunes of each company.  Of
course, the diversification benefits of the pool depends critically on how
correlated the component securities are to each other. Assuming that
mortgages in a given pool are statistically independent leads to
substantial diversification benefits, and greatly simplifies the
bond-pricing analytics, but may not accurately reflect reality.

Not surprisingly, aggregating the risks, expected returns, and
correlations of a collection of mortgages to price a given ABS-CDO is
accomplished through a mathematical formula.  One of the most widely used
formulas is the so-called ``Gaussian copula'' developed by Li (2000),
which is a specific functional form that is both analytically tractable
and sufficiently general to capture the joint statistical properties of a
large number of correlated mortgages.%
\footnote{Despite the impact of Li's (2000) formula on industry practice,
the literature on credit models is considerably broader, hence the
negative publicity surrounding Li's contribution may be misplaced. See
Duffie and Singleton (2003) and Caouette et al.\ (2008) for a more
complete review of this literature.}
In a February 23, 2009 article in \textit{Wired\/} by Salmon (2009) titled
``Recipe for Disaster: The Formula That Killed Wall Street'', this
contribution was described in the following manner:
\begin{quote}
\baselineskip=14pt%
One result of the collapse has been the end of financial economics as
something to be celebrated rather than feared. And Li's Gaussian copula
formula will go down in history as instrumental in causing the
unfathomable losses that brought the world financial system to its knees.
\end{quote}
According to this account, the financial crisis was caused by a single 
mathematical expression. 

However, in attempting to expose the specific shortcomings of this model, 
Salmon contradicts his story's captivating title and premise, 
inadvertently providing a more nuanced explanation for the crisis: 
\begin{quote}
\baselineskip=14pt%
The damage was foreseeable and, in fact, foreseen. In 1998, before Li had
even invented his copula function, Paul Wilmott wrote that ``the
correlations between financial quantities are notoriously unstable.''\ \
Wilmott, a quantitative-finance consultant and lecturer, argued that no
theory should be built on such unpredictable parameters. And he wasn't
alone. During the boom years, everybody could reel off reasons why the
Gaussian copula function wasn't perfect. Li's approach made no allowance
for unpredictability: It assumed that correlation was a constant rather
than something mercurial. Investment banks would regularly phone
Stanford's Duffie and ask him to come in and talk to them about exactly
what Li's copula was. Every time, he would warn them that it was not
suitable for use in risk management or valuation.

In hindsight, ignoring those warnings looks foolhardy. But at the time, it
was easy. Banks dismissed them, partly because the managers empowered to
apply the brakes didn't understand the arguments between various arms of
the quant universe. Besides, they were making too much money to stop.

Li's copula function was used to price hundreds of billions of dollars'
worth of CDOs filled with mortgages. And because the copula function used
CDS prices to calculate correlation, it was forced to confine itself to
looking at the period of time when those credit default swaps had been in
existence: less than a decade, a period when house prices soared.
Naturally, default correlations were very low in those years. But when the
mortgage boom ended abruptly and home values started falling across the
country, correlations soared.
\end{quote}
\noindent%
Apparently, the combination of historically low correlation of residential
mortgage defaults and the profitability of the CDO business caused
managers to ignore the warnings issued by a number of quants.  But Salmon
(2009) makes an even bolder claim:
\begin{quote}
\baselineskip=14pt%
Bankers securitizing mortgages knew that their models were highly
sensitive to house-price appreciation. If it ever turned negative on a
national scale, a lot of bonds that had been rated triple-A, or risk-free,
by copula-powered computer models would blow up. But no one was willing to
stop the creation of CDOs, and the big investment banks happily kept on
building more, drawing their correlation data from a period when real
estate only went up.
\end{quote}
If this claim is true, the natural follow-up question is why did no one 
allow for the possibility of a national-level housing market decline in 
their models? 

\begin{figure}[htbp]
\centering
\includegraphics[clip, trim=0 190 0 -175, scale=.45]%
{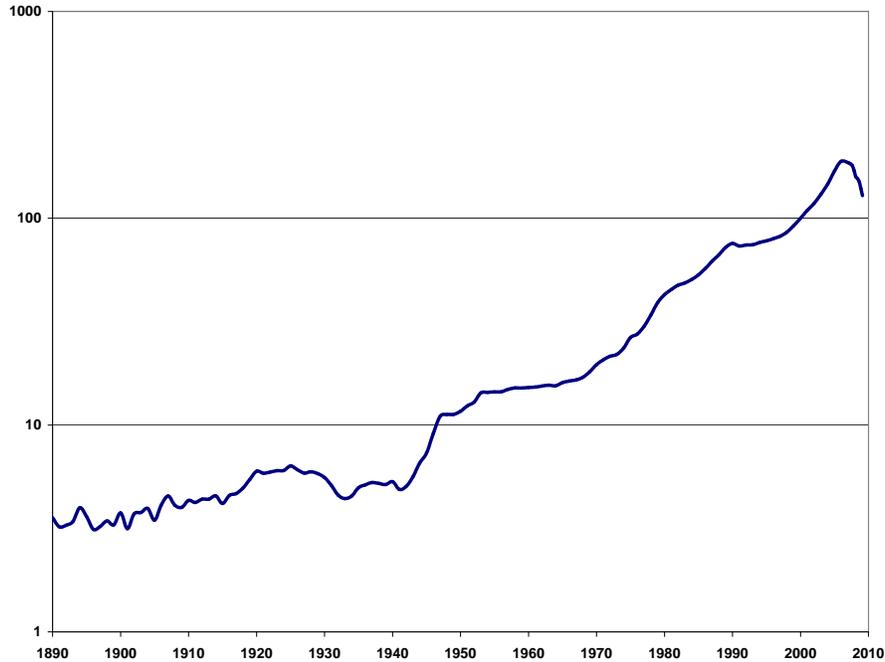}%
\caption{Nominal U.S.\ residential home price index from 1890 to 2009:Q1,
on a logarithmic scale.
Source:  Robert J.\ Shiller.}
\label{fig:homeprices}
\end{figure}

One answer may be found in Figure \ref{fig:homeprices}, which contains a
plot of the national U.S. residential nominal home price index constructed
by Yale economist Robert J.\ Shiller from 1890 to 2009Q2.%
\footnote{The vertical axis is scaled logarithmically so that a given 
percentage change in home prices corresponds to the same vertical 
increment regardless of what year it occurred and the level of the index 
at that time.} 
This striking graph shows that the U.S.\ has not experienced a significant
national housing-market downturn since 1933.  And as long as housing
prices do not decline, it is easy to see why the default correlations of a
geographically diversified pool of mortgages will remain low.  On the
other hand, once housing prices do decline on a national level, it is also
easy to see why default correlations might spike up abruptly.  Contrary to
Salmon's (2009) claim that mortgage default correlation estimates were
biased by the shorter estimation window when CDS spreads were available
and home prices were rising, Figure \ref{fig:homeprices} suggests that
using historical data from the previous seven decades would not have
materially changed the outcome.

This chain of logic brings us to the main point of this example.  If, over
a 70-year period, the U.S.\ did not experience a significant
national-level decline, how could the possibility of such a decline be
factored into the analysis? From our perspective, there is a two-part
answer.

First, it must be acknowledged that the pricing of ABS-CDOs involves a
significant degree of Level-4 uncertainty, implying that assumptions of
stationarity and fixed laws of motion do not apply. Therefore, no single
formula will yield a complete description of the relevant risks and
rewards of any security, no matter how sophisticated. Because of the
presence of Level-4 uncertainty, the need for additional institutional and
economic structure is considerably greater.

One example of such additional structure is provided by the post-crisis
analysis of Khandani, Lo, and Merton (2009) in which they argue that the
heightened default correlations during housing-market declines are due to
the combination of low interest rates, rising home prices, and greater
access to cheap refinancing opportunities that typically precede declines.
Individually, each of the three conditions is benign, but when they occur
simultaneously, as they did over the past decade, they impose an
unintentional synchronization of homeowner leverage.  This
synchronization, coupled with the indivisibility of residential real
estate that prevents homeowners from selling a fraction of their homes to
reduce leverage when property values decline and homeowner equity
deteriorates, conspire to create a ``ratchet'' effect in which homeowner
leverage is maintained or increased during good times without the ability
to decrease leverage during bad times.  If refinancing-facilitated
homeowner-equity extraction is sufficiently widespread---as it was during
the years leading up to the peak of the U.S.\ residential real-estate
market in June 2006---the inadvertent coordination of leverage during a
market rise implies higher correlation of defaults during a market drop.

While this analysis also uses quantitative methods---in particular,
numerical simulation and derivatives pricing models---the methods are
simple in comparison to the institutional and economic structure on which
the simulations are based.  By incorporating richer and more accurate
institutional features, the proportion of irreducible uncertainty in CDO
pricing models can be reduced.

Second, the ultimate response to Level-4 uncertainty is human judgment, 
and Salmon's (2009) observation that senior management had neither the 
motives nor the required expertise to limit their risk exposures to the 
CDO market is, ultimately, a behavioral argument.  During extended periods 
of prosperity, the individual and collective perception of risk declines, 
e.g., as losses become historically more remote, human perception of small 
probabilities of loss quickly converge to zero (Lichtenstein et 
al., 1982).%
\footnote{If readers have any doubt about this human tendency, they should 
ask long-time residents of California how they live with the constant 
threat of earthquakes.} 
Although a national decline in home prices was certainly possible,
historical experience suggested it was highly improbable. Therefore, it is
not surprising that during the period from 1998 to 2006, senior managers
at major financial institutions were not concerned with this risk---the
last such event was too far back for anyone to care about, especially
given how much financial markets had changed in the interim.

In the case of CDOs, another relevant aspect of human behavior is the
possibility that a number of decisionmakers simply did not have the
technical expertise to properly evaluate the risk/reward trade-offs of
these securities.  As Salmon (2009) puts it:
\begin{quote}
\baselineskip=14pt%
Bankers should have noted that very small changes in their underlying
assumptions could result in very large changes in the correlation number.
They also should have noticed that the results they were seeing were much
less volatile than they should have been---which implied that the risk was
being moved elsewhere. Where had the risk gone?

They didn't know, or didn't ask. One reason was that the outputs came from
``black box'' computer models and were hard to subject to a commonsense
smell test. Another was that the quants, who should have been more aware
of the copula's weaknesses, weren't the ones making the big
asset-allocation decisions. Their managers, who made the actual calls,
lacked the math skills to understand what the models were doing or how
they worked. They could, however, understand something as simple as a
single correlation number. That was the problem.
\end{quote}

Quantitative illiteracy is not acceptable in science.  Although financial 
economics may never answer to the same standards as physics, nevertheless, 
managers in positions of responsibility should no longer be allowed to 
take perverse anti-intellectual pride in being quantitatively illiterate 
in the models and methods on which their businesses depend.

\subsection{Too Many Quants, or Not Enough?}
A third example of misguided reaction to the financial crisis is the 
impression that there were too many quants on Wall Street and their 
concerted efforts created the crisis.  We find this response to the crisis 
particularly troubling because it is inconsistent with the equally popular 
view that the crisis was caused by securities too complex for anyone to 
fully understand.  If anything, the preliminary evidence accumulated from 
the aftermath of the crisis suggests that we need more 
financial-engineering expertise and training at the most senior levels of 
management at banks, broker-dealers, insurance companies, mutual funds, 
pension funds, and regulatory agencies, not less. The excess demand for 
financial expertise has only increased over time as markets have become 
more complex, competitive, and globally integrated. 

A skeptic might respond by claiming that it may be preferable to return to
a simpler time with fewer complex securities, alleviating the need for
quantitatively sophisticated stakeholders.  However, this view misses the
fact that risk pooling and risk transfer have been essential components of
capital markets for centuries, with innovations that have arisen in
response to the legitimate needs of individuals and corporations.%
\footnote{Merton's (1992, 1993) functional approach to financial
innovation suggests that new products and services may vary considerably
over time, but the functions they serve are more stable.  See also Crane
et al.\ (1995).}
As with most new technologies, problems arise when their limits are not
fully understood or when they are purposely abused to attain short-run
benefits at the expense of safety or stability.  But when properly modeled
and managed, securitization can play an extremely positive role in the
risk-pooling and risk-transfer functions that capital markets are uniquely
designed to accomplish.  This potential has no doubt contributed to the
growing popularity of quants in the financial industry.

\begin{figure}[htbp]
\centering
\includegraphics[clip, trim=0 225 0 -150, scale=.6]%
{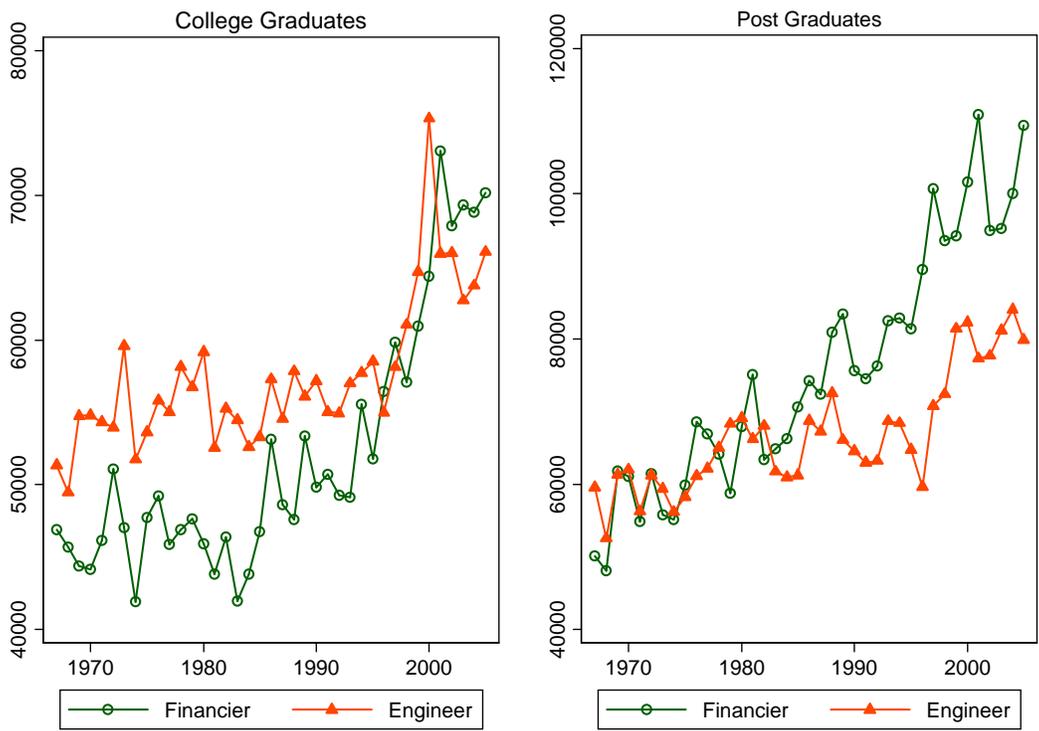}%
\caption{Comparison of finance and engineering graduates from 1967 to
2005, in 2001 U.S. dollars.  Source:  Philippon and Reshef (2007).}
\label{fig:salaries}
\end{figure}

Indirect evidence for an excess demand of finance expertise is documented
in Philippon and Reshef's  (2007) comparison of the annual incomes of
U.S.\ engineers and finance-trained graduates from 1967 to 2005.  The
comparison between finance and engineering students is a useful one
because both are technical disciplines, and over the past 20 years,
engineers have been making significant inroads into the finance labor
market.  Figure \ref{fig:salaries} shows that until the mid-1980's,
college graduates in engineering enjoyed higher incomes than college
graduates in finance, and post-graduates in engineering had about the same
compensation as post-graduates in finance.  However, since the 1980's,
finance-trained college graduates have caught up to their engineering
counterparts, and surpassed them in 2000 and every year thereafter.  But
the more impressive comparison is for post-graduates-since 1982, the
annual income of finance post-graduates has exceeded that of engineers
every year, and the gap has widened steadily over these two decades.  This
pattern suggests that the demand for financial expertise has grown
considerably during this time.

Table \ref{tbl:degrees}, which reports the number of MIT engineering and
finance degrees awarded from 1999 to 2007, provides another perspective on
the dearth of financial expertise.  In 2007, MIT's School of Engineering
graduated 337 Ph.D.'s in engineering; in contrast, the MIT Sloan School of
Management produced only 4 finance Ph.D.s.  These figures are not unique
to MIT, but are, in fact, typical among the top engineering and business
schools. Now, it can be argued that the main focus of the Sloan School is
its M.B.A. program, which graduates approximately 300 students each year,
but most M.B.A. students at Sloan and other top business schools do not
have the technical background to implement models such as Li's (2009)
Gaussian copula formula for ABSs and CDOs, nor does the standard M.B.A.\
curriculum include courses that cover such models in any depth. Such
material---which requires advanced training in arcane subjects such as
stochastic processes, stochastic calculus, and partial differential
equations---is usually geared towards Ph.D.\ students.%
\footnote{However, due to the growth of the derivatives business over the
past decade, a number of universities have begun to offer specialized
Master's-level degree programs in financial engineering and mathematical
finance to meet the growing demand for more technically advanced students
trained in finance. Whether or not such students are sufficiently prepared
to fill the current knowledge gap in financial technology remains to be
seen.}

\begin{table}[htbp]
\centering
\includegraphics[clip, trim=0 600 0 0, scale=.8]%
{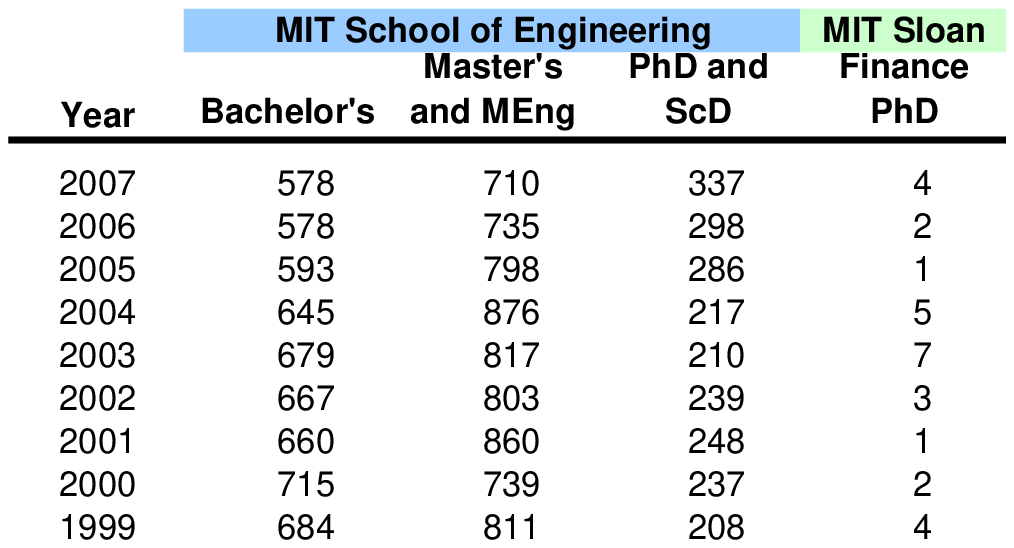}%
\caption{Number of MIT degrees awarded in engineering and in finance from
1999 to 2007.  Source: MIT Annual Report of the President, 1999 to 2007.}
\label{tbl:degrees}
\end{table}

The disparity between the number of Ph.D.s awarded in engineering and 
finance in Table \ref{tbl:degrees} raises the question of why such a 
difference exists. One possible explanation may be the sources of funding. 
MIT engineering Ph.D.\ students are funded largely through government 
grants (DARPA, DOE, NIH, and NSF), whereas MIT Sloan Ph.D.\ students are 
funded exclusively through internal MIT funds.  Given the importance of 
finance expertise, one proposal for regulatory reform is to provide 
comparable levels of government funding to support finance Ph.D.\ 
students.  Alternatively, funding for finance Ph.D. students might be 
raised by imposing a small surcharge on certain types of derivatives 
contracts, e.g., those that are particularly complex or illiquid and, 
therefore, contribute to systemic risk.  This surcharge may be viewed as a 
means of correcting some of the externalities associated with the impact 
of derivatives on systemic risk.  A minuscule surcharge on, say, credit 
default swaps, could support enough finance Ph.D. students at every major 
university to have a noticeable and permanent impact on the level of 
financial expertise in both industry and government. 

\section{Conclusion}
\label{sec:conclusion}
Financial economics may be a long way from physics, but this state of 
affairs is cause for neither castigation nor celebration---it is merely a 
reflection of the dynamic, non-stationary, and ultimately human aspect of 
economic interactions.  In sharp contrast to the other social sciences, 
economics does exhibit an enviable degree of consistency across its many 
models and methods.  In the same way that scientific principles are 
compact distillations of much more complex phenomena, economic principles 
such as supply-and-demand, general equilibrium, the no-arbitrage 
condition, risk/reward trade-offs, and Black-Scholes/Merton 
derivatives-pricing theory capture an expansive range of economic 
phenomena. However, any virtue can become a vice when taken to an extreme, 
particularly when that extreme ignores the limitations imposed by 
uncertainty.

In this respect, the state of economics may be closer to disciplines such 
as evolutionary biology, ecology, and meteorology.  In fact, weather 
forecasting has become noticeably better in our lifetimes, no doubt 
because of accumulated research breakthroughs and technological advances 
in that field.  However, like economics, it is impossible to conduct 
experiments in meteorology, even though the underlying physical principles 
are well understood.  Despite the fact that long periods of ``typical'' 
weather allow us to build simple and effective statistical models for 
predicting rainfall, on occasion a hurricane strikes and we are reminded 
of the limits of those models (see Masters, 1995).  And for the truly 
global challenges such as climate change, the degree of subjectivity and 
uncertainty gives rise to spirited debate, disagreement, and what appears 
to be chaos to uninformed outsiders. Should we respond by discarding all 
forecasting models for predicting rainfall, or should we simply ignore the 
existence of hurricanes because they fall outside our models? 

Perhaps a more productive response is to delineate the domain of validity 
of each model, to incorporate this information into every aspect of our 
activities, to attempt to limit our exposure to the catastrophic events 
that we know will happen but which we cannot predict, and to continue 
developing better models through data collection, analysis, testing, and 
reflection, i.e., becoming smarter. 

We all use models of one sort or another---with wildly varying degrees of 
reliability---in making sense of the past and attempting to look into the 
future.  Mental models are virtually synonymous with human intelligence 
(see, for example, Gigerenzer, 2000, and Hawkins, 2004), and some models 
are explicit, systematic, and quantitative while others are more intuitive 
or discretionary, based on ineffable expertise. Qualitative information is 
difficult to incorporate in the former, while testability, repeatability, 
and scalability are challenges to the latter. And both approaches are, in 
the financial industry, routinely and intentionally hidden behind a 
proprietary veil, further hampering a direct assessment of their 
respective costs and benefits. 

Faith in any person or organization claiming to have a deep and intuitive
grasp of market opportunities and risks is no better or worse than putting
the same faith and money behind a mysterious black-box strategy.  What
matters in each case is the transparency of the process, an opportunity to
assess the plausibility and limitations of the ideas on which a strategy
is based, clarity about expectations for risks as well as returns, an
alignment of incentives between the investment manager and the investor,
and proper accountability for successes and failures.  Part of the recent
distrust of quantitative models is that they are inanimate, and therefore
inherently difficult to trust. Even when we manage to develop a certain
level of comfort with a quantitative strategy, ongoing due diligence is
needed to assess the consistency, integrity, and incentives of the human
beings responsible for implementing the strategy. It is important to
distinguish our emotional needs from the formal process of assessing the
models themselves, for which a degree of quantitative literacy will always
be required.

So what does this imply for the future of finance?  Our hope is that the 
future will be even brighter because of the vulnerabilities that the 
recent crisis has revealed.  By acknowledging that financial challenges 
cannot always be resolved with more sophisticated mathematics, and 
incorporating fear and greed into models and risk-management protocols 
explicitly rather than assuming them away, we believe that the financial 
models of the future will be considerably more successful, even if less 
mathematically elegant and tractable. Just as biologists and 
meteorologists have broken new ground thanks to computational advances 
that have spurred new theories, we anticipate the next financial 
renaissance to lie at the intersection of theory, practice, and 
computation. 

While physicists have historically been inspired by mathematical elegance
and driven by pure logic, they also rely on the ongoing dialogue between
theoretical ideals and experimental evidence.  This rational, incremental,
and sometimes painstaking debate between idealized quantitative models and
harsh empirical realities has led to many breakthroughs in physics, and
provides a clear guide for the role and limitations of quantitative
methods in financial markets, and the future of finance.
\newpage

\center{\large\bf References} \addcontentsline{toc}{section}{References}
\normalsize \baselineskip 12pt \vskip 30pt

\begin{description}
\item
Arrow, Kenneth, 1964, ``The Role of Securities in the Optimal
Allocation of Risk Bearing'', \textit{Review of Economic Studies} 31,
91--96.


\item
Blinder, A., 2009, ``Six Errors on the Path to the Financial Crisis'',
\textit{New York Times}, January 24.

\item
Bouchaud, J., Farmer, D. and F. Lillo, 2009, ``How Markets Slowly 
Digest Changes in Supply and Demand'', in H. Thorsten and K. 
Schenk-Hoppe, ed.s, \textit{Handbook of Financial Markets: Dynamics 
and Evolution}.  Amsterdam: Elsevier. 

\item
Brennan, T. and A. Lo, 2009, ``The Origin of Behavior'', available at
SSRN: \\ {\tt http://ssrn.com/abstract=1506264}.


\item
Caouette, J., Altman, E., Narayanan, P. and R. Nimmo, 2008,
\textit{Managing Credit Risk: The Great Challenge for Global Financial
Markets}, 2nd edition.  New York: Wiley Finance.

\item
Coffee, J., 2008, ``Analyzing the Credit Crisis: Was the SEC 
Missing in Action?'', {\it New York Law Journal}, December 5. 

\item
Cootner, P., ed., 1964, \textit{The Random Character of Stock Market
Prices}.  London, Risk Books.

\item
Crane, D., Froot, K., Mason, S., Perold, A., Merton, R., Bodie, Z.,
Sirri, E. and P. Tufano, 1995, \textit{The Global Financial System: A
Functional Perspective}.  Boston, MA: Harvard Business School Press.

\item Damasio, A., 1994, \textit{Descartes' Error: Emotion,
Reason, and the Human Brain}.  New York: Avon Books.


\item
Debreu, G., 1959, \textit{Theory of Value}.  New York: John Wiley and
Sons.

\item
Debreu, G., 1991, ``The Mathematization of Economics'',
\textit{American Economic Review} 81, 1--7.

\item
de Jong, T. and W. van Soldt, 1989, ``The Earliest Known Solar
Eclipse Record Redated'', \textit{Nature} 338, 238--240

\item
Diaconis, P., Holmes, S. and R. Montgomery, 2007, ``Dynamical Bias in
the Coin Toss'', \textit{SIAM Review} 49, 211--235.

\item
Duffie, D. and K. Singleton, 2003, \textit{Credit Risk: Pricing,
Measurement, and Management}.  Princeton, NJ: Princeton University
Press.


\item
Elster, J., 1998, ``Emotions and Economic Theory'', \textit{Journal of
Economic Literature} 36, 47--74.

\item
Farmer, D., 2002, ``Market Force, Ecology and Evolution'', 
\textit{Industrial and Corporate Change} 11, 895--953. 

\item
Farmer, D. and A. Lo, 1999, ``Frontiers of Finance: Evolution and 
Efficient Markets'', {\it Proceedings of the National Academy of 
Sciences} 96, 9991--9992. 

\item
General Accounting Office, 1999, \textit{Long-Term Capital Management: 
Regulators Need to Focus Greater Attention on Systemic Risk}, 
GAO/GGD--00--3.  Washington, D.C.: U.S.\ General Accounting 
Office.

\item
Government Accountability Office, 2009, \textit{Financial Markets 
REgulation: Financial Crisis Highlights Need to Improve Oversight of 
Leverage at Financial Institutions and across System}, GAO--09--739.  
Washington, D.C.: U.S.\ Government Accountability Office. 

\item
Gigerenzer, G., 2000, \textit{Adaptive Thinking: Rationality in the
Real World}.  Oxford: Oxford University Press.

\item
Gitterman, M., 2005 \textit{The Noisy Oscillator: The First Hundred 
Years, From Einstein to Now}. Singapore: World Scientific Publishing. 

\item
Grossberg, S. and W. Gutowski, 1987, ``Neural Dynamics of Decision
Making Under Risk: Affective Balance and Cognitive-Emotional
Interactions'', \textit{Psychological Review} 94, 300--318.

\item
Hamilton, J., 1989, ``A New Approach to the Economic Analysis of
Nonstationary Time Series and the Business Cycle'',
\textit{Econometrica} 57, 357--384.

\item
Hamilton, J., 1990, ``Analysis of Time Series Subject to Changes in
Regime'', \textit{Journal of Econometrics} 45, 39--70.

\item
Hawkins, J., 2004, \textit{On Intelligence}.  New York: Henry Holt and
Company.

\item
Hull, A. and J. White, 2009, ``The Risk of Tranches Created from
Residential Mortgages'', Working Paper, Joseph L.\ Rotman School of
Management, University of Toronto, May.

\item
Khandani, A. and A. Lo, 2007, ``What Happened to the Quants in August
2007?'', \textit{Journal of Investment Management} 5, 5--54.

\item
Khandani, A. and A. Lo, 2008, ``What Happened To The Quants In August
2007?: Evidence from Factors and Transactions Data'', available at
SSRN:\\ {\tt http://ssrn.com/abstract=1288988}.

\item
Klein, L., 1970, \textit{An Essay on the Theory of Economic
Prediction}.  Chicago, IL: Markham Publishing Company.

\item
Knight, F., 1921, \textit{Risk, Uncertainty, and Profit}.  Boston, MA:
Houghton Mifflin.

\item
Labaton, S., 2008, ``The Reckoning'', {\it New York Times}, October 3. 

\item Lehmann, B., 1990, ``Fads, Martingales, and Market Efficiency'',
\textit{Quarterly Journal of Economics} 105, 1--28.

\item
Li, D., 2000, ``On Default Correlation: A Copula Function Approach'',
\textit{Journal of Fixed Income} 9, 43--54.


\item
Lintner, J., 1965, ``The Valuation of Risky Assets and the Selection
of Risky Investments in Stock Portfolios and Capital Budgets'',
\textit{Review of Economics and Statistics} 47, 13--37.

\item
Lloyd, C., 2008, \textit{What on Earth Happened?: The Complete Story
of the Planet, Life, and People from the Big Bang to the Present Day}.
New York: Bloomsbury USA.

\item Lo, A., 1999, ``The Three P's of Total Risk Management'', \textit{Financial
Analysts Journal} 55, 13--26.


\item
Lo, A., 2004, ``The Adaptive Markets Hypothesis: Market Efficiency
from an Evolutionary Perspective'', {\it Journal of Portfolio
Management} 30, 15--29.

\item Lo, A.~and C.~MacKinlay, 1990a, ``When Are Contrarian Profits Due to
Stock Market Overreaction?'', {\it Review of Financial Studies} 3,
175--206.

\item
Lo, A. and C. MacKinlay, 1990b, ``Data Snooping Biases in Tests of
Financial Asset Pricing Models'', \textit{Review of Financial Studies}
3, 431--468.

\item
Lo, A. and C. MacKinlay, 1999, \textit{A Non-Random Walk Down Wall
Street}.  Princeton, NJ: Princeton University Press.

\item
Lo, A. and R. Merton, 2009, ``Preface to the \textit{Annual Review of 
Financial Economics}'', \textit{Annual Review of Financial Economics} 
1, 1--18. 

\item
Lo, A. and D. Repin, 2002, ``The Psychophysiology of Real-Time
Financial Risk Processing'', {\it Journal of Cognitive Neuroscience}
14, 323--339.


\item
Loewenstein, G., 2000, ``Emotions in Economic Theory and Economic
Behavior'', \textit{American Economic Review} 90, 426--432.

\item
Lorenz, E., 1963, ``Deterministic Nonperiodic Flow'', \textit{Journal
of Atmospheric Science} 20, 130--141.

\item
Lucas, R., 1972, ``Expectations and the Neutrality of Money'',
\textit{Journal of Economic Theory} 4, 103--124.


\item
Mantegna, R. and E. Stanley, 1994, ``Stochastic Process with
Ultra-Slow Convergence to a Gaussian: The Truncated Levy Flight'',
\textit{Physical Review Letters} 73, 2946--2949.

\item
Mantegna, R. and E. Stanley, 2000, \textit{An Introduction to
Econophysics: Correlations and Complexity in Finance}.  Cambridge, UK:
Cambridge University Press.

\item
Masters, T., 1995 \textit{Neural, Novel \& Hybrid Algorithms for Time 
Series Prediction}.  New York: John Wiley and Sons. 

\item
Merton, R., 1992, ``Financial Innovation and Economic Performance'',
\textit{Journal of Applied Corporate Finance} 4, 12--22.

\item
Merton, R., 1993, ``Operation and Regulation in Financial
Intermediation: A Functional Perspective'', in P.\ Englund, ed.,
\textit{Operation and Regulation of Financial Markets}. Stockholm: The
Economic Council.

\item
Muth, J., 1961, ``Rational Expectations and the Theory of Price
Movements'', \textit{Econometrica} 29, 315--335.

\item
Nash, J., 1951, ``Non-Cooperative Games'', \textit{Annals of
Mathematics} 54, 286--295.

\item
Peters, E. and P. Slovic, 2000, ``The Springs of Action: Affective and
Analytical Information Processing in Choice'', \textit{Personality and
Social Psychology Bulletin} 26, 1465--1475.

\item
Pickard, L., 2008, ``Viewpoint: SEC's Old Capital Approach was 
Tried---and True'', {\it American Banker}, August 8. 

\item
Philippon, T. and A. Reshef, 2007, ``Skill Biased Financial
Development: Education, Wages and Occupations in the U.S. Financial
Sector'', Working Paper, Stern School of Business, New York
University.

\item
Rabin, M., 1998, ``Psychology and Economics'', \textit{Journal of 
Economic Literature} 36, 11--46.

\item
Rabin, M., 2002, ``A Perspective on Psychology and Economics'',
\textit{European Economic Review} 46, 657--685.

\item
Reif, F., \textit{Fundamentals of Statistical and Thermal Physics}.
New York: McGraw Hill.




\item
Rolls, E., 1999, \textit{The Brain and Emotion}.  Oxford, UK: Oxford
University Press.

\item
Samuelson, P., 1947, \textit{Foundations of Economic Analysis}.
Cambridge, MA: Harvard University Press.

\item
Samuelson, P., 1998, ``How Foundations Came to Be'', \textit{Journal
of Economic Literature} 36, 1375--1386.

\item
Samuelson, P., 2009, ``An Enjoyable Life Puzzling over Modern Finance 
Theory'', \textit{Annual Review of Financial Economics} 1, 19--36. 

\item
Salmon, F., 2009, ``Recipe for Disaster: The Formula That Killed Wall
Street'', \textit{Wired}, February 23.

\item
Satow, J., 2008, ``Ex-SEC Official Blames Agency for Blow-Up of 
Broker-Dealers'', \textit{New York Sun}, September 18. 

\item
Scarf, H. (with the collaboration of T.\ Hansen), 1973, \textit{The
Computation of General Equilibrium}.  New Haven, CT: Yale University
Press.

\item
Sharpe, W. , 1964, ``Capital Asset Prices: A Theory of Market
Equilibrium under Conditions of Risk'', \textit{Journal of Finance}
19, 425--442.

\item
Sirri, E., 2009 ``Securities Markets and Regulatory Reform'', speech
delivered to the National Economists Club, April 9, available at \\
{\tt http://www.sec.gov/news/speech/2009/spch040909ers.htm}.

\item
Solow, R., 1956, ``A Contribution to the Theory of Economic Growth'',
\textit{Quarterly Journal of Economics} 70, 65--94.

\item
Stiglitz, J., 1987, ``Principal and Agent'', \textit{The New Palgrave:
A Dictionary of Economics} 3, 966-971.

\item
Stiglitz, J., 2009, ``Capitalist Fools'', \textit{Vanity Fair}, 
January. 

\item
Strogatz, Steven H., 1994, \textit{Nonlinear Systems and Chaos}.  New
York: Perseus Publishing.

\item
Taleb, N., 2007, \textit{The Black Swan: The Impact of the Highly 
Improbable}.  New York: Random House. 

\item
Theil, H., 1971, \textit{Principles of Econometrics}.  New York: John
Wiley \& Sons.

\item
Tinbergen, J., 1956, \textit{Economic Policy: Principles and Design}.
Amsterdam: North-Holland Publishing Company.

\item
Tobin, J., 1958, ``Liquidity Preference as Behavior Towards Risk'',
\textit{Review of Economic Studies} 67, 65--86.

\item
von Neumann, J. and O. Morgenstern, 1944, \textit{Theory of Games and
Economic Behavior}.  Princeton, NJ: Princeton University Press.

\item
Weinberg, S., 1977, \textit{The First Three Minutes}.  New York:
Bantam Books.

\item
White, H., 1984, \textit{Asymptotic Theory for Econometricians}.  New
YorK: Academic Press.

\item
Woodward, S., 2009, ``The Subprime Crisis of 2008: A Brief Background
and a Question'', ASSA Session on Recent Financial Crises, January 3,
2009.


\end{description}
\end{document}